\documentclass{article} 
\usepackage[T1]{fontenc}
\usepackage[utf8]{inputenc}
\usepackage[margin=3.5cm]{geometry}
\usepackage{amssymb,amsmath,bm,natbib,xcolor,hyperref}
\usepackage{authblk}
 \usepackage{grffile}
\usepackage{graphicx,subfigure}
\usepackage{pifont}
\usepackage{soul}
\definecolor{verde}{rgb}{0,.56,0}

\title{\bf Centrifugal instability of Stokes layers in crossflow: \\
	the case of a forced cylinder wake}
\author[1]{\sc Juan D'Adamo}
\author[2]{\sc Ramiro Godoy-Diana}
\author[2]{\sc Jos\'e Eduardo Wesfreid}
\affil[1]{\normalsize Facultad de Ingenier\'ia Universidad de Buenos Aires (CONICET), Av.
	Paseo Col\'on 850, C1063ACV - Buenos Aires - Argentina.}
\affil[2]{\normalsize Physique et M\'ecanique des Milieux Het\'erog\`enes (PMMH), CNRS UMR
	7636; ESPCI ParisTech; UPMC; Univ. Paris Diderot (Paris 7), 10 rue
	Vauquelin, 75005 Paris, France}

\date{}

\begin{document}
\maketitle

\begin{abstract}
The wake flow around a circular cylinder at $Re\approx100$ performing
rotatory oscillations has been thoroughly discussed in the literature, mostly 
focusing on the modifications to the natural B\'enard-von K\'arm\'an vortex
street that result from the forced shedding modes locked to the rotatory
oscillation frequency. The usual experimental and theoretical frameworks at
these Reynolds numbers are quasi-two-dimensional, since the secondary
instabilities bringing a three-dimensional structure to the cylinder wake flow
occur only at higher Reynolds numbers. In the present paper we show that a
three-dimensional structure can appear below the usual three-dimensionalization
threshold, when forcing with frequencies lower than the natural vortex shedding
frequency, at high amplitudes, as a result of a previously unreported mechanism:
a pulsed centrifugal instability of the oscillating Stokes layer at the wall of
the cylinder. The present numerical investigation lets us in this way propose a
physical explanation for the turbulence-like features reported in the recent
experimental study of \cite{dadamo2011b}.
\end{abstract}
\maketitle
\section{Introduction}
A circular cylinder performing rotational oscillations around its axis in an
infinite viscous fluid produces an 
axisymmetric pulsed boundary layer, called a Stokes layer. This is a flow
susceptible to generate centrifugal 
instabilities. The linear stability problem of this flow configuration has been 
studied by \cite{hall1975,seminara1976} using asymptotic methods. A threshold for the appearance of three-dimensional (3D) axisymmetric instability modes was determined. \cite{riley1976} did also stability calculations  not directly on the  Stokes layer problem but considering the modulated circular Couette flow under axisymmetric disturbances, in the narrow-gap limit.
 Later, \cite{aouidef1994, ern1998,ern1999, ern2002} considered this flow as a limit case for the stability problem of the classic 
geometry of two concentric cylinders with oscillation: the Taylor-Couette 
configuration (see e.g. 
\cite{chandrasekhar1981} for a review). In both 
cases, 
the control parameter is the Taylor number, defined as 
\begin{equation}\label{taylor1}
T=R_i  \sqrt{\frac{d}{\mathcal R}}
\end{equation}
where $R_i=\omega_i r_i d /\nu$ is a Reynolds number based on the rotational
angular velocity of the cylinder $\omega_i$. We keep the notation of the
Taylor-Couette configuration, where the subscript $i$ stands for \emph{inner}
cylinder, $r_i$ thus being the radius of the cylinder. In addition, $\mathcal R$
is the local radius of curvature and $\nu$ the 
kinematic viscosity. The characteristic length scale $d$ in the Taylor-Couette
case is the gap between the cylinders, which fixes the scale of the wavelength
of the primary instability.  For the case studied by \cite{seminara1976},
however, the instability occurs in the inner Stokes boundary layer of thickness
$\delta_S=\sqrt{\nu/\omega_i}$ around the oscillating cylinder. They have 
determined analytically, numerically and experimentally the critical values for
$T$  associated with the onset of a 
Taylor-Couette-type vortex flow. Vortices evenly spaced, with a critical 
length  $\lambda_c$ in the cylinder axial direction which is proportional to 
$\delta_S$, are thus developed.
\begin{figure}[t]
\centering
\includegraphics[width=0.75\textwidth,trim=0mm 0mm 0mm 0mm,clip]{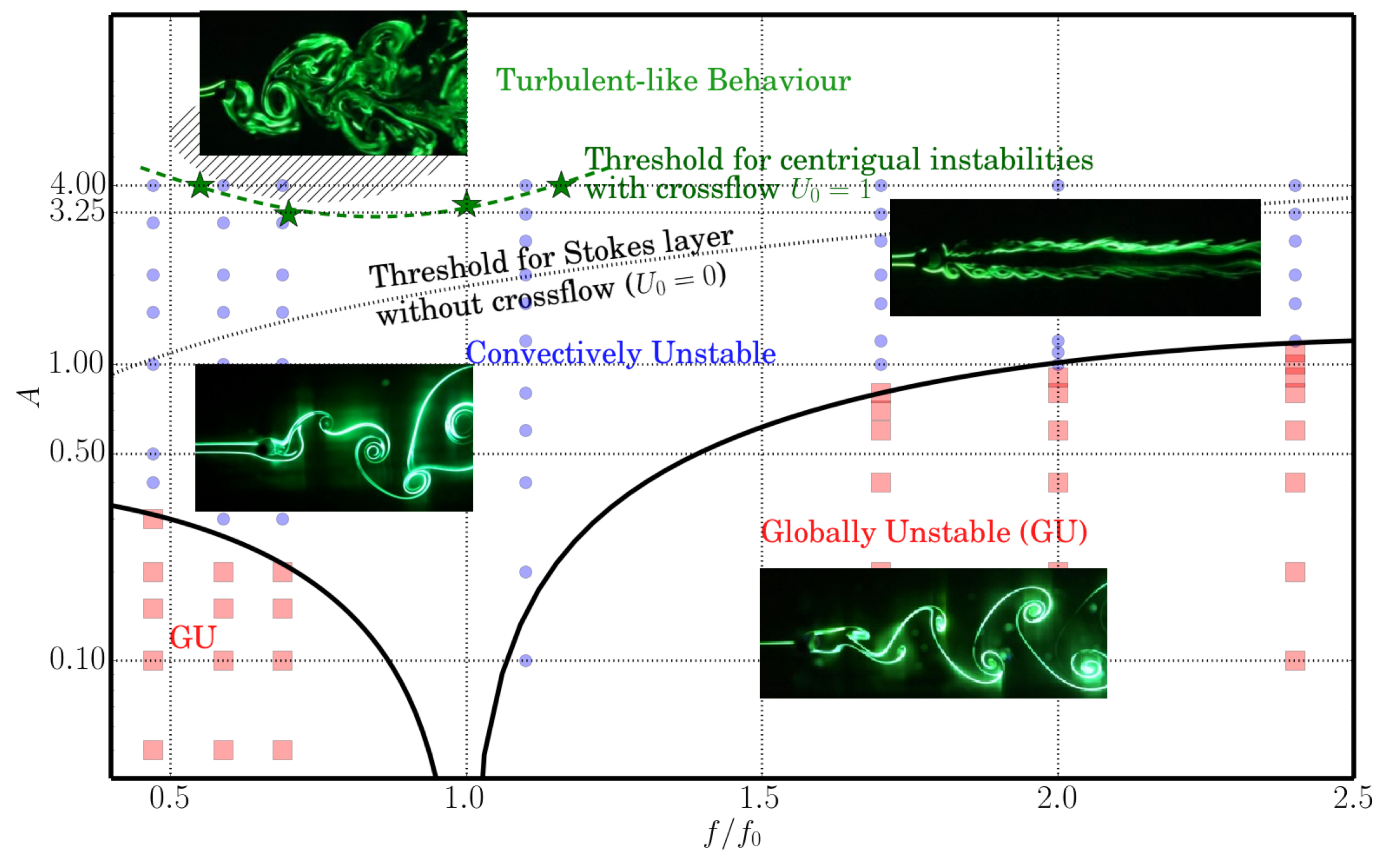}             
\caption{Different flow states for the forcing 
parameters $(f,A)$ scrutinised in \cite{dadamo2011b}. Visualisations are from 
\cite{thiria2006}. Solid 
lines represent the threshold from global to convective instability.  The 
dotted line indicates the threshold to centrifugal instability of the Stokes 
layer of an oscillating cylinder without a crossflow given by $T_c=165$, from 
\cite{seminara1976}. Symbols \textcolor{verde}{$\bigstar$} show the threshold 
for 3D centrifugal instabilities observed in the 3D DNS discussed in Figs.
\ref{scale_fig1-c} and \ref{scale_fig1-d}. Dashed region stands for 
states with 
turbulent-like behaviour described in \cite{dadamo2011b}. }\label{introfig}
\end{figure}

On the other hand, when a uniform flow comes across a cylinder,  a prototypical 
2D wake flow takes place for moderate free-stream Reynolds numbers 
$Re=DU_0/\nu$,
 where $D$ is the diameter of the cylinder, $U_0$ the free-stream velocity. The 
 well-known B\'enard-von K\'arm\'an (BvK) vortex street 
\cite[][]{benard1908,vonkarman1911} results from the destabilisation of the
steady flow in the wake of the cylinder and produces the periodic shedding of
opposite-signed vortices with a frequency $f_0$, that occurs above the 
threshold $Re_c\approx 47$ (see e.g. \cite{provansal1987,jackson1987}). This
flow is quasi-two dimensional up to $Re\lesssim 180$.
In a recent work \cite[][]{dadamo2011b}, we studied experimentally the problem 
of the forced wake performing rotary oscillations at $Re=100$.  The rotational
oscillation 
of the cylinder is prescribed by a forcing function of frequency $f$ 
and amplitude $\theta_0$ that can be written as $\theta(t)=\theta_0\cos(2\pi f
t)$, 
which allows the forcing to be unequivocally described using two independent 
non-dimensional parameters as did by \cite{taneda1978}: the forcing amplitude 
$A=u_{\theta\mathrm{max}}/U_0$, where $u_{\theta\mathrm{max}}=D\pi f\theta_0$ is
the 
maximal azimuthal velocity of the rotational oscillation, and the ratio $f/f_0$.
We characterised the spatial development of the flow and its stability 
properties following previous studies by \cite{thiria2006,thiria2007}. A
synthesis of the 
case study is presented in Figure \ref{introfig}. From the analysis of power
density spectra of the flow we gave a 
detailed description of the forced wake, giving insight on the energy 
distribution, the different frequency components, and in particular on a 
continuous spectrum observed for a high amplitude of the forcing oscillation. 
Furthermore, vortex structures revealed turbulence-like features like splitting 
and mixing in a spatial cascade pattern. A question remained concerning the
physical mechanism 
present in the bifurcation that triggers such behaviour of the wake. 

We speculated on a 3D centrifugal instability to be at the origin of this 
sequence 
of transitions. A natural
first attempt to test this idea is shown in figure \ref{introfig}, where the 
critical Taylor number $T_c=165$ corresponding to the instability threshold of 
the pure rotatory oscillating cylinder case without crossflow studied by 
\cite{seminara1976} is 
identified in the frequency-amplitude phase space $(f,A)$ of the forced wake of 
\cite{dadamo2011b}. This crude estimate for a threshold is compatible with the 
experimental points where the turbulent-like behaviour was observed (low 
frequencies and high amplitudes of the forcing oscillation). The purpose of the 
present paper is to characterize in detail the existence of a 3D instability and
its centrifugal nature, using analytical estimations from the 2D flow and from
3D direct numerical 
simulations (DNS).

It is worth mentioning that centrifugal instabilities were also reported for 
forced flows with different 
configurations. For transverse oscillations of a cylinder in a fluid at rest, 
\cite{honji1981}, obtained visualizations that identified 3D structures
produced by centrifugal instabilities. \cite{hall1984} performed a stability analysis
of this configuration and gave a theoretical explanation. \cite{tatsuno1990}
 investigated in detail the patterns and the structure of the flows that result from these instabilities.
Later, \cite{elston2004} addressed DNS calculations and  Floquet stability analysis for this problem.

Three dimensional instabilities in wake flows have been studied theoretically
and numerically by \cite{blackburn2005} where it was determined 
that bifurcations to three-dimensionality  can occur from a 
two-dimensional time-periodic base state with space–time
reflection symmetry for the wake of symmetrical bluff bodies. More recently for
the case 
of the two-dimensional stationary flow past a rotating cylinder,
\cite{pralits2013}
suggest that the stationary unstable three-dimensional mode could be the result
of a hyperbolic instability.

\cite{lojacono2010} were interested on the role of rotationally oscillations can modify the three-dimensional transition in the wake of a cylinder. The frequency of oscillation was matched to the natural vortex-shedding frequency, $f^+=1$, for Re = 300. They reported changes on the three dimensional modes from Floquet stability analysis on two-dimensional periodic flow. They found that the rotational oscillation dramatically suppressed mode B, even for small amplitudes of oscillation. Mode A was also damped, but not as significantly as mode B. For what they considered high rotational oscillation amplitudes, in our notation $A\simeq0.66$ they identify a new three-dimensional transition mode, which they called D mode, that shares the same symmetries as mode A.

Three dimensional characteristics of forced wakes have been recently studied by \cite{kumar2013} for the case of rotational oscillations at $Re=185$ near the transition, using flow visualization, hot-wire anemometry and
PIV. Spatial distribution of lock-on regions and its relationship with the forcing frequencies and amplitudes were determined. They also found that for certain forcing parameters $(f^+, A)$, the flow can be forced to become two-dimensional. Studied amplitudes were up to $A=\pi$, a value below the threshold found in \cite{dadamo2011b} by means of spatio-temporal spectral analysis.

To summarise, the present work sets up a new view about 3D instabilities in
wake 
flows, which have often been discussed in the case of the circular cylinder for 
$Re>180$  as secondary instabilities to the BvK vortex street. 
We organise the paper as follows: in the next section we describe the method
used for 
the direct numerical simulations (DNS); results are 
presented in Section \ref{resultados} where we determine the 
3D stability threshold; in Section \ref{theory_back} we investigate the
instability 
nature, using some concepts of centrifugal instabilities and propose therefore a
reduction of the complex problem; lastly we elaborate 
our conclusions in Section \ref{conclusiones}, showing 
analogies with the Taylor-Couette problem of eccentric cylinders.

\section{Problem definition for DNS}
In order to study this problem, we performed 2D and 3D direct numerical 
simulations with Gerris free software,  
a parallelised tree-based adaptive solver for Navier-Stokes 
equations \cite{Popinet2003572}. The code  combines an  adaptive multi-grid 
finite volume method and the methods of immersed boundary and volume of fluid 
(VOF). The basic equations   are the incompressible continuity equation and 
Navier-Stokes equations, which can be written in terms of the velocity
$\mathbf{u}=(u,v,w)$ and pressure $p$ fields as:
\begin{eqnarray}
 \nabla\cdot \bm u=0\nonumber\\
 \frac{\partial \bm u}{\partial t}+(\bm u \cdot\nabla)\bm u=-\frac{\nabla 
p}{\rho}+\nu\nabla^2\bm u 
\end{eqnarray}

\begin{figure}[t]
{\centering
\subfigure[]{\label{problem-a}
\includegraphics[height=25mm,trim=5mm 2mm 2mm 2mm 2mm,clip]{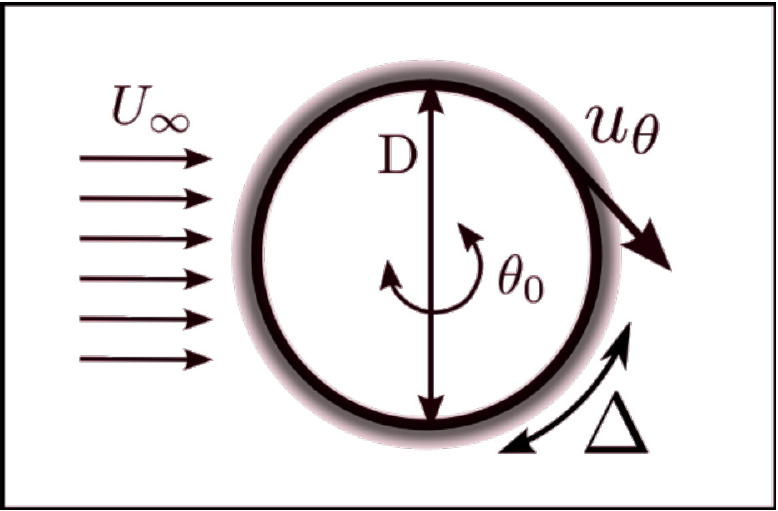}}
\subfigure[]{\label{problem-b}
\includegraphics[height=30mm,trim=40mm 39mm 30mm 30mm,clip]{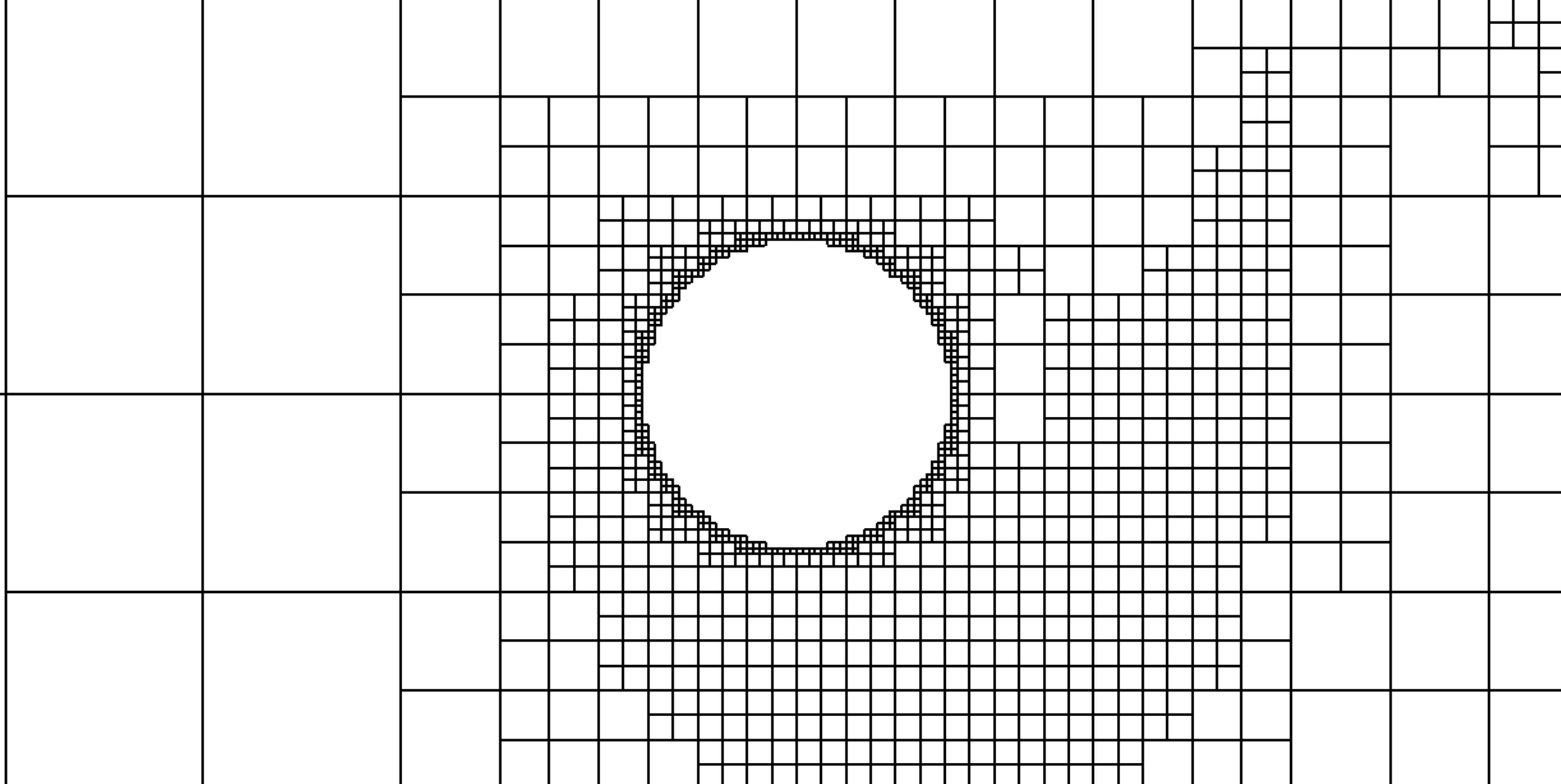}}\\
\subfigure[]{\label{problem-c}
\includegraphics[width=.7\textwidth,trim=0mm 0mm 0mm 0mm 0mm,clip]{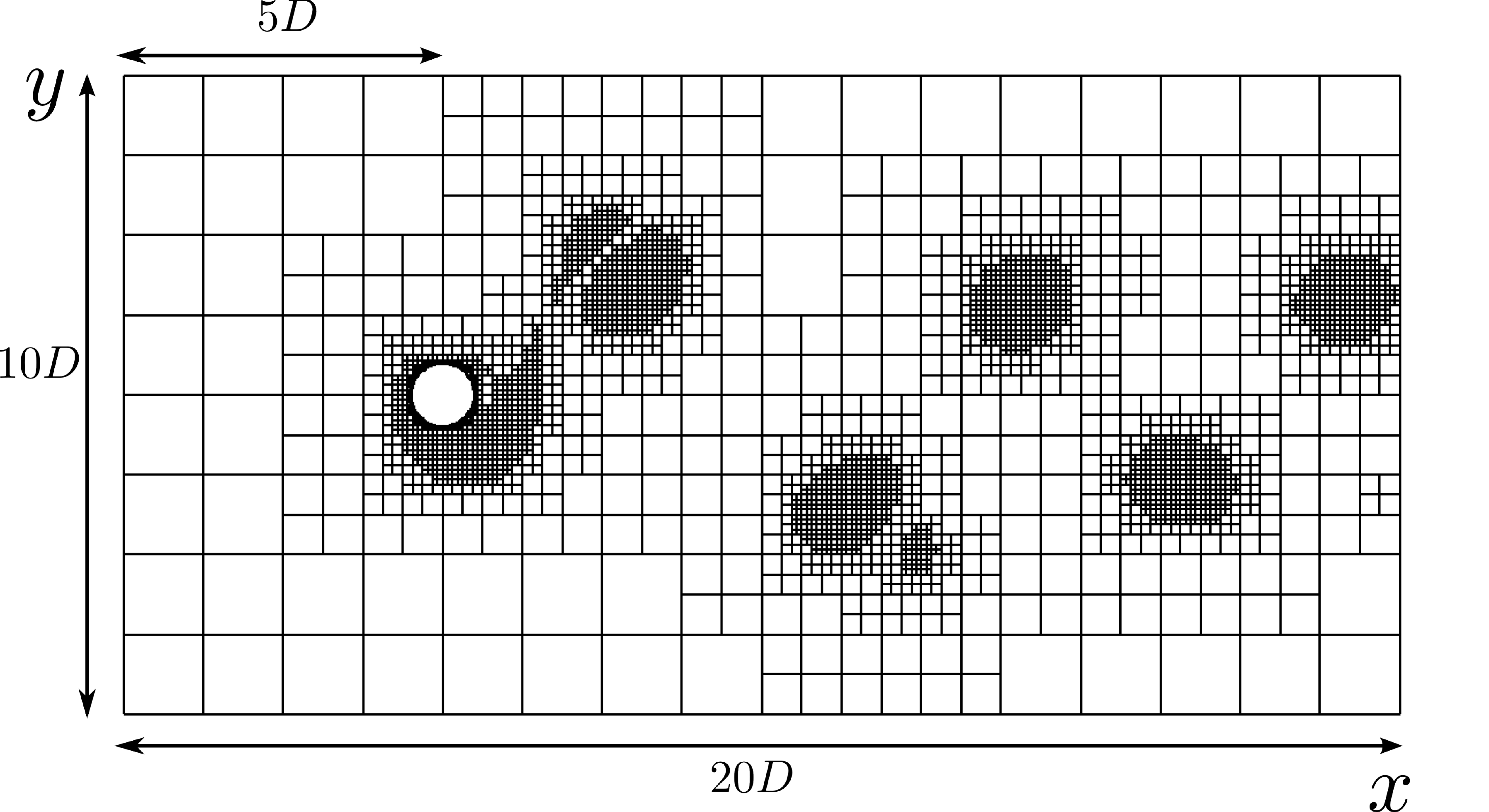}
\put(-315,100){$L_y=$}
\put(-187,5){$L_x=$}
}
\caption{a) Parameters of the problem. b) The adaptive 2D mesh around the
circular 
cylinder shows the octree 
structure. c) Problem domain.}}\label{problem}
\end{figure}
The domain is spatially discretised using cubic finite volumes organised 
hierarchically as an octree.  Along with the forcing problem parameters, a 2D 
example of the spatial discretisation is given in Figure \ref{problem}.  The 
flow
domain, shown in Figure \ref{problem-c} is $L_x\times L_y=20D\times10D$ for 2D 
simulations 
and  $L_z=20D$ for the spanwise direction in 3D 
simulations. As detailed in \cite{Popinet2003572} the mesh can be refined near 
the solid boundary and it can  use vorticity gradients as an adaptive 
criterion. A cell 
is refined whenever:
\begin{equation}
 \frac{|\nabla\times \bm u|\Delta x}{\max|\bm u|}>\xi
\end{equation}
where $\Delta x$ is the size of the cell and $\xi$ is a user-defined 
threshold which can be interpreted as the maximum angular deviation (caused by 
the local vorticity) of a particle travelling at speed $\max|\bm u|$. This
adaptive criterion is represented in Figure \ref{problem-b} and \ref{problem-c}
 where different box sizes are noticeable. In order to reveal BvK vortices as 
well 
as centrifugal structures, we 
choose a  minimum grid size of $D/51.2$ for the solid boundary and $D/12.8$ to 
define vortex regions. The $\xi$ threshold is set to 0.05 for 3D 
simulations and to 0.01 for 2D simulations. 
The flow parameters of the simulations are defined in order to match the
experimental case of \cite{dadamo2011b}: Cross flow velocity $U_\infty=1$,
kinematic viscosity $\nu=10^{-3}$ and cylinder diameter $D=0.1$, giving a 
Reynolds number $Re=100$. \\

The boundary conditions are: $u=1$ for $x=-5D$; 
$u=1$ for $y=\pm 5$; the outflow condition is $\partial v/\partial x=0$ and 
$p=0$ for $x=15D$; for 3D simulations,  a symmetry condition is used for the 
flow at $z=20D$; and at the cylinder surface, 
$\mathbf{u}=\mathbf{u}_{solid}$ where $\mathbf{u}_{solid}$ depends on the 
forcing.  As depicted in Figure \ref{problem-a} rotatory oscillations are 
characterised by an angular coordinate $\theta(t)=\theta_0\cos(\alpha)$, where
the 
forcing phase is $\alpha=2\pi f_f t$, and 
tangential displacements $\Delta=u_{\theta}/(2\pi f_f)$. Given $f_0$ the  
natural frequency of vortex shedding, the forcing frequency $f_f$ is 
written in dimensionless form as $f^+=f_f / f_0$. A non-dimensional number for
the amplitude of 
oscillations is obtained by comparing the maximum tangential velocity 
$u_{\theta\mathrm{max}}$ and the free flow velocity, 
$A=u_{\theta\mathrm{max}}/U_\infty$.

\section{Results of the numerical simulation}\label{resultados}
\begin{figure}\begin{center}
\subfigure[]{\label{resul1-a}
 \includegraphics[width=.47\textwidth,trim=0mm 0mm 0mm 
0mm,clip]{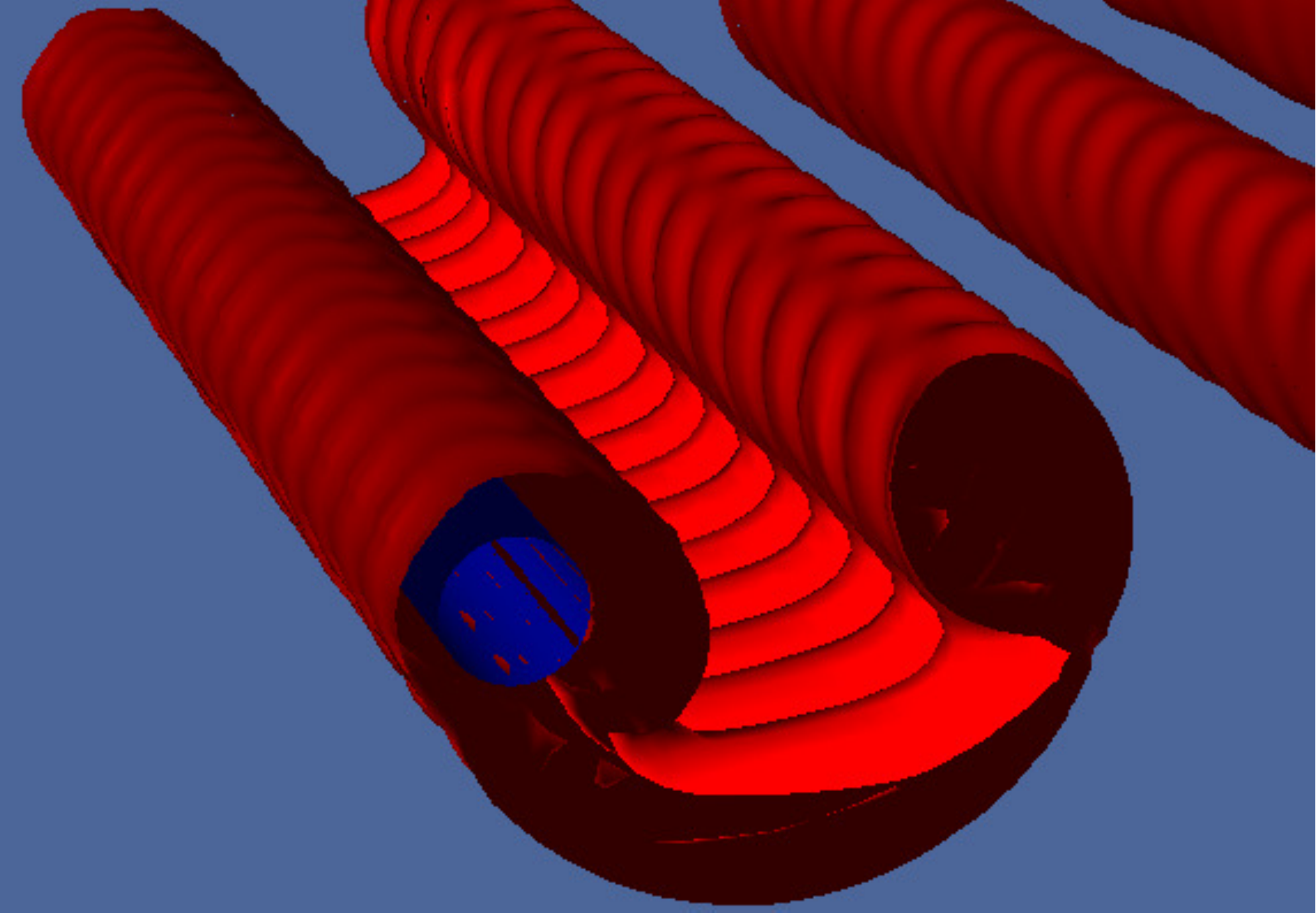}}
\subfigure[]{\label{resul1-b}
 \includegraphics[width=.47\textwidth,trim=0mm 10mm 0mm 
0mm,clip]{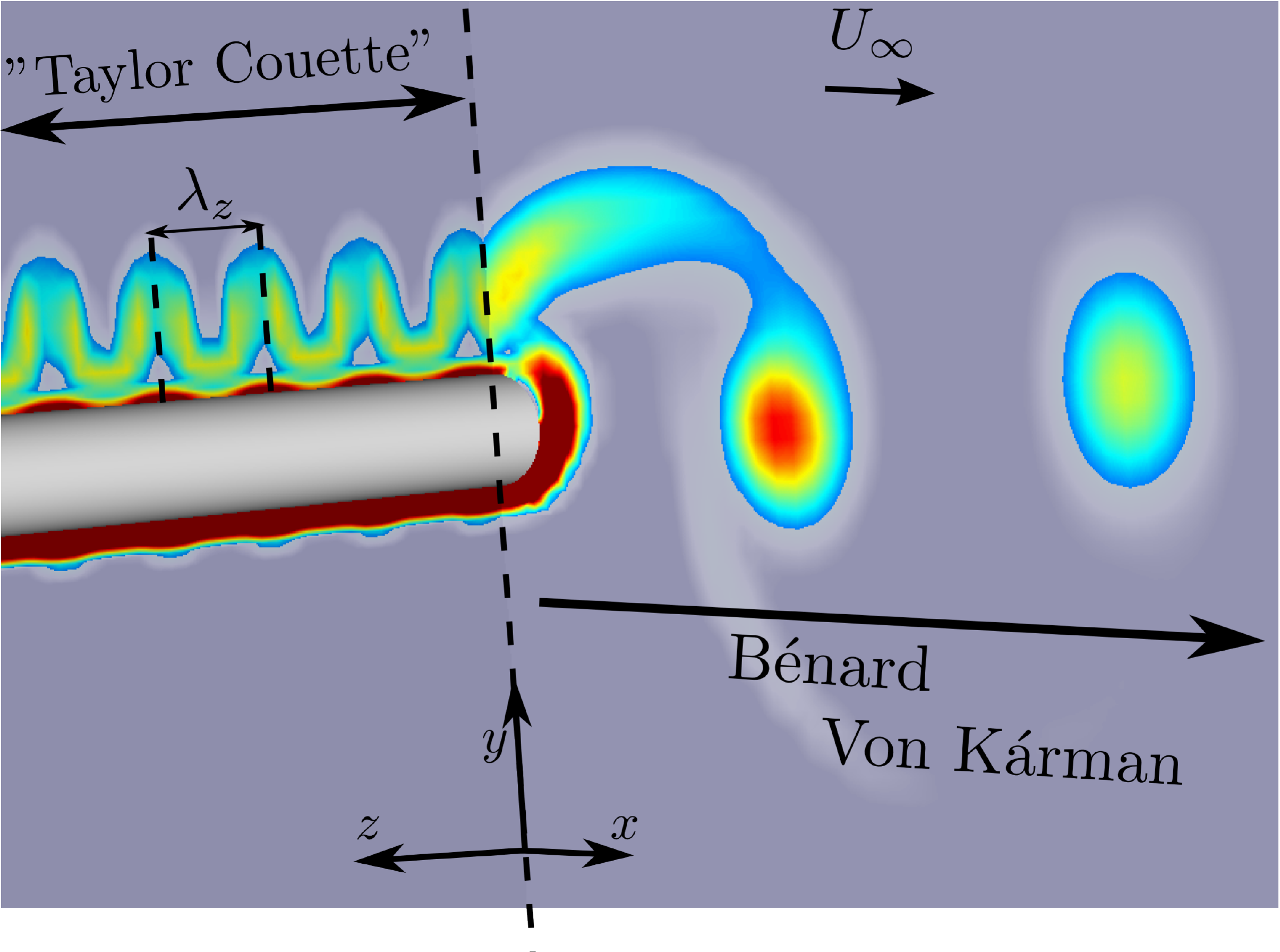}
}
 \caption{a) Isosurface for total vorticity modulus $\|\bar\omega\|$ for 
$f^+=0.75$, $A=4.00$. b) Vorticity modulus contours for two cutting planes 
revealing the main flow structures ($f^+=0.75$, $A=4.00$). }\label{resul1}
\end{center}
\end{figure}
\begin{figure}\begin{center}
		\subfigure[]{\label{resul2-a}
			\includegraphics[width=.47\textwidth,trim=0mm 0mm 0mm 
			0mm,clip]{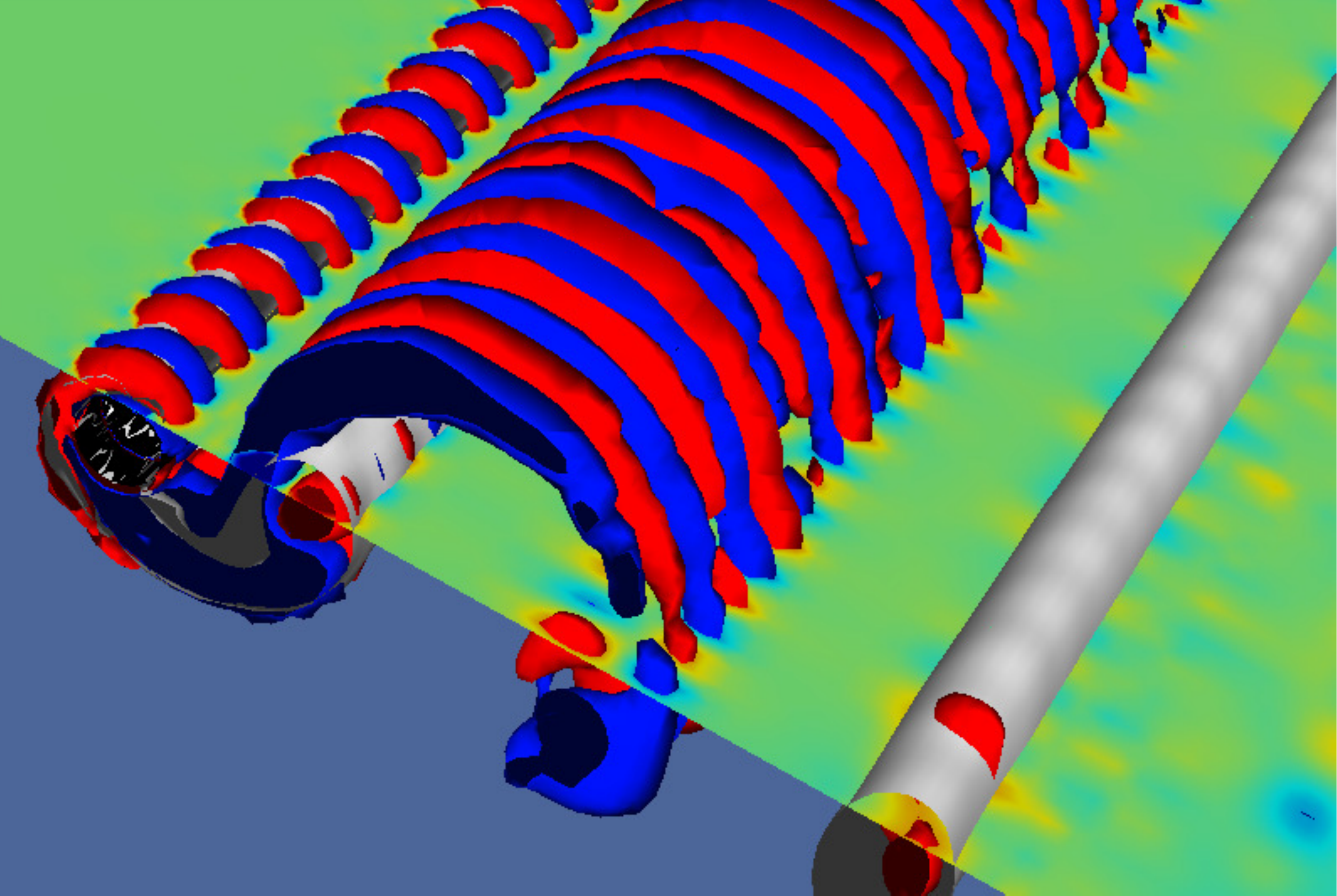}}
		\subfigure[]{\label{resul2-b}
			\includegraphics[width=.47\textwidth,trim=0mm 53mm 0mm 
			0mm,clip]{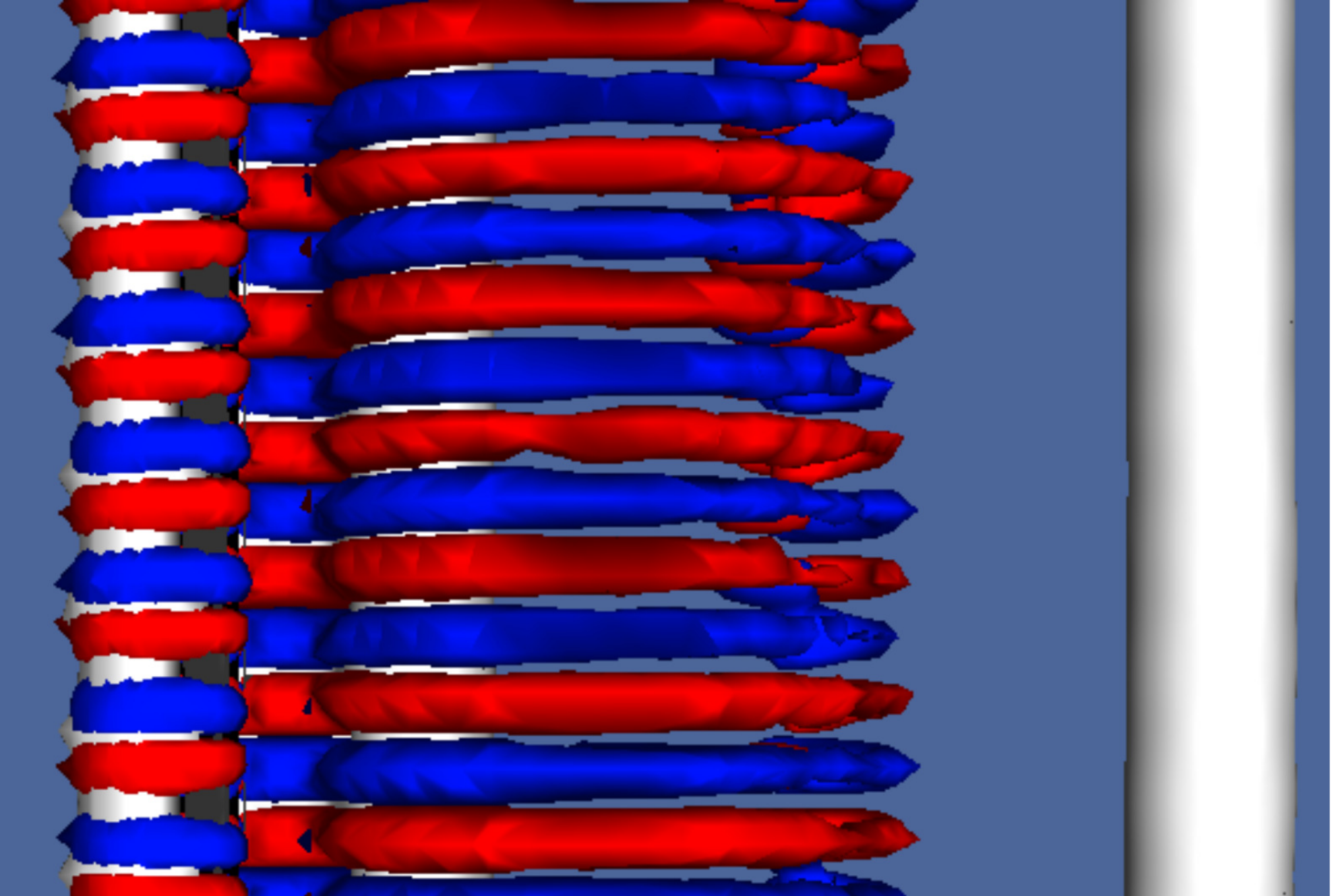}
		}
		\caption{For forcing parameters $f^+=0.75$, $A=4.00$. a) Isosurfaces for streamwise vorticity $\omega_x$: $\omega_x=3$ (red), $\omega_x=-3$ (blue). The white isosurface represents spanwise vorticity $\omega_z=15$
			 b) Top view for isosurfaces for streamwise and spanwise vorticity. }\label{resul2}
	\end{center}
\end{figure}

We first performed 3D DNS numerical simulations. Figure \ref{resul1} shows a
case with the  forcing parameters $(f^+=0.75,A=4.00)$.
The isosurface of vorticity modulus in Figure \ref{resul1-a} shows on one side 
the classic BvK wake structure synchronised with the forcing frequency. 
Additionally, a previously not reported effect is also clear: the modulation of
the vorticity field along the direction of the cylinder axis. The two effects
are depicted in Figure 
\ref{resul1-b}, revealing the 3D vortex structure around the cylinder and a
well-defined wavelength $\lambda_z$. Moreover, Figure \ref{resul2} shows the spatial distribution of $\omega_x$ along with $\omega_z$ for $f^+=0.75$, $A=4.00$, which allows us to consider the symmetry properties of the observed mode.
	The spatio-temporal symmetry, $H$ , of the two-dimensional flow is defined as:
	\begin{equation}\label{eqsymmetry}
		H \omega(x,t)=K_y\omega(x,t+T/2)=(-\omega_x,\omega_y,-\omega_z)(x,-y,t+T/2)
	\end{equation}
	where $K_y$ is a spatial reflection. For an $H$-symmetric flow, from \ref{eqsymmetry}, the $x$-vorticity changes sign with $t\rightarrow t + T /2$ and $y \rightarrow -y$ at any fixed $(x, z)$. This is the case for mode A, whereas for mode B, the sign of x-vorticity does not change.
	
	As studied by \cite{blackburn2005}, there are exactly three codimension-one bifurcations from a two-dimensional time-periodic base state to three-dimensional flow that are observable with
	variations in a single parameter. In this regard, \cite{lojacono2010} showed that oscillatory forcing at $Re=300$ leads to the appearance of a different mode (mode D) which has the same symmetries of mode A. Considering the symmetries observed in the present case (Figure \ref{resul2}), the identified structures are not $H$-invariant, and they share the same symmetry as mode B.

In what follows we thoroughly scrutinise the onset of this 3D pulsed instability.
\begin{figure}\begin{center}
\subfigure[]{\label{scale_fig1-a}
 \includegraphics[width=.6\textwidth]{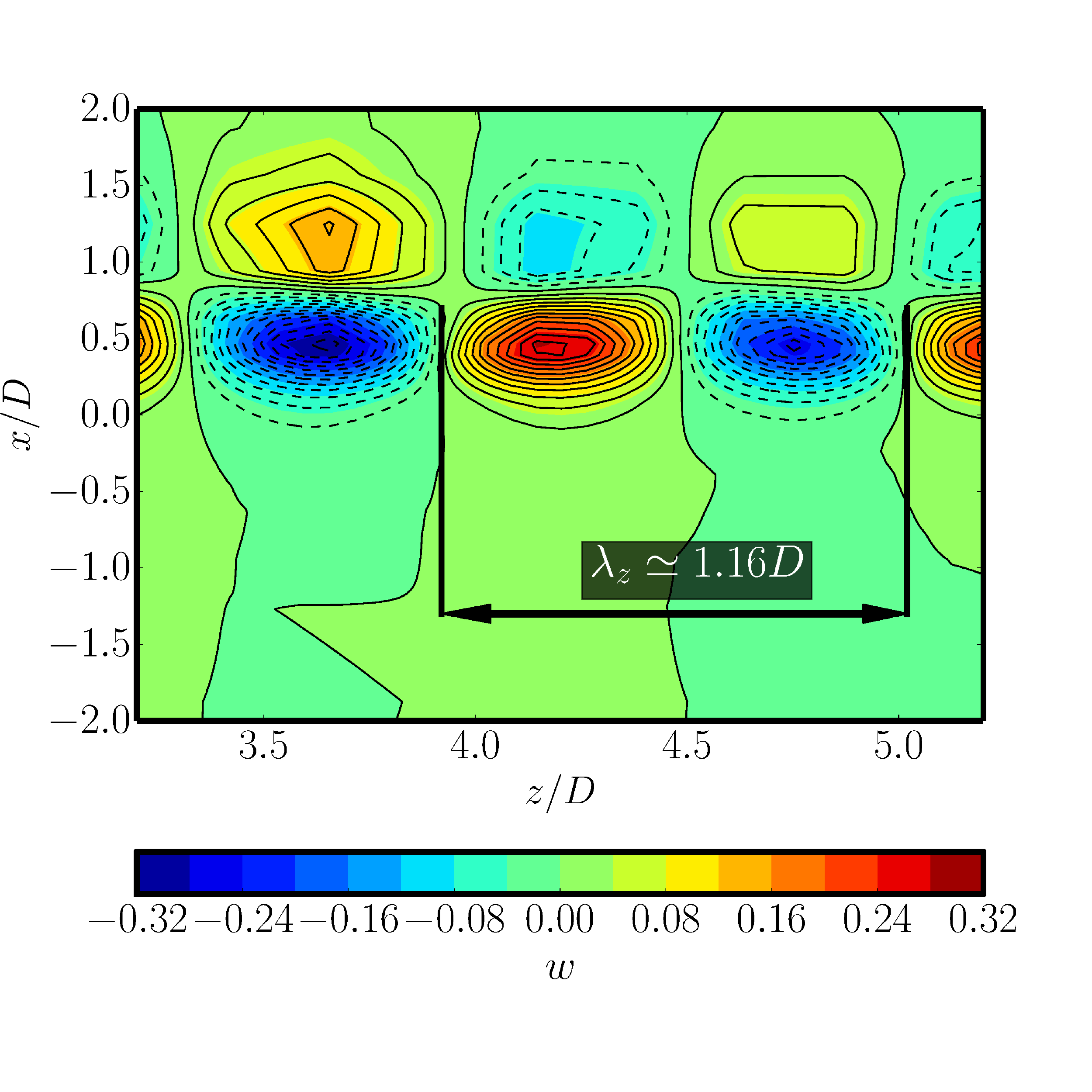}}\\
 \subfigure[]{\label{scale_fig1-c}
 \includegraphics[height=.29\textheight,trim=0mm 0mm 0mm 
0mm,clip]{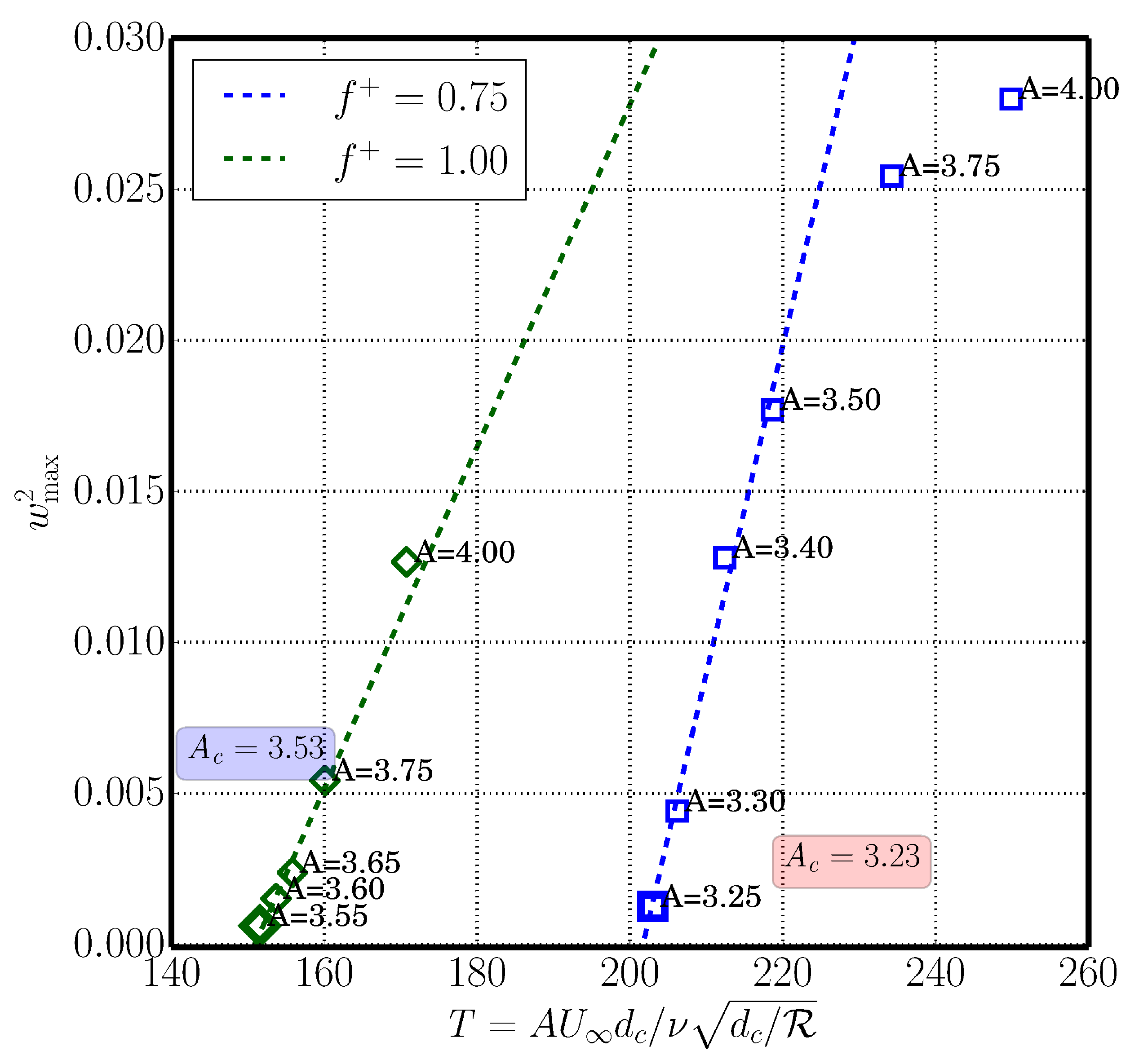}}
 \subfigure[]{\label{scale_fig1-d}
 \includegraphics[height=.29\textheight,trim=0mm 0mm 0mm 
0mm,clip]{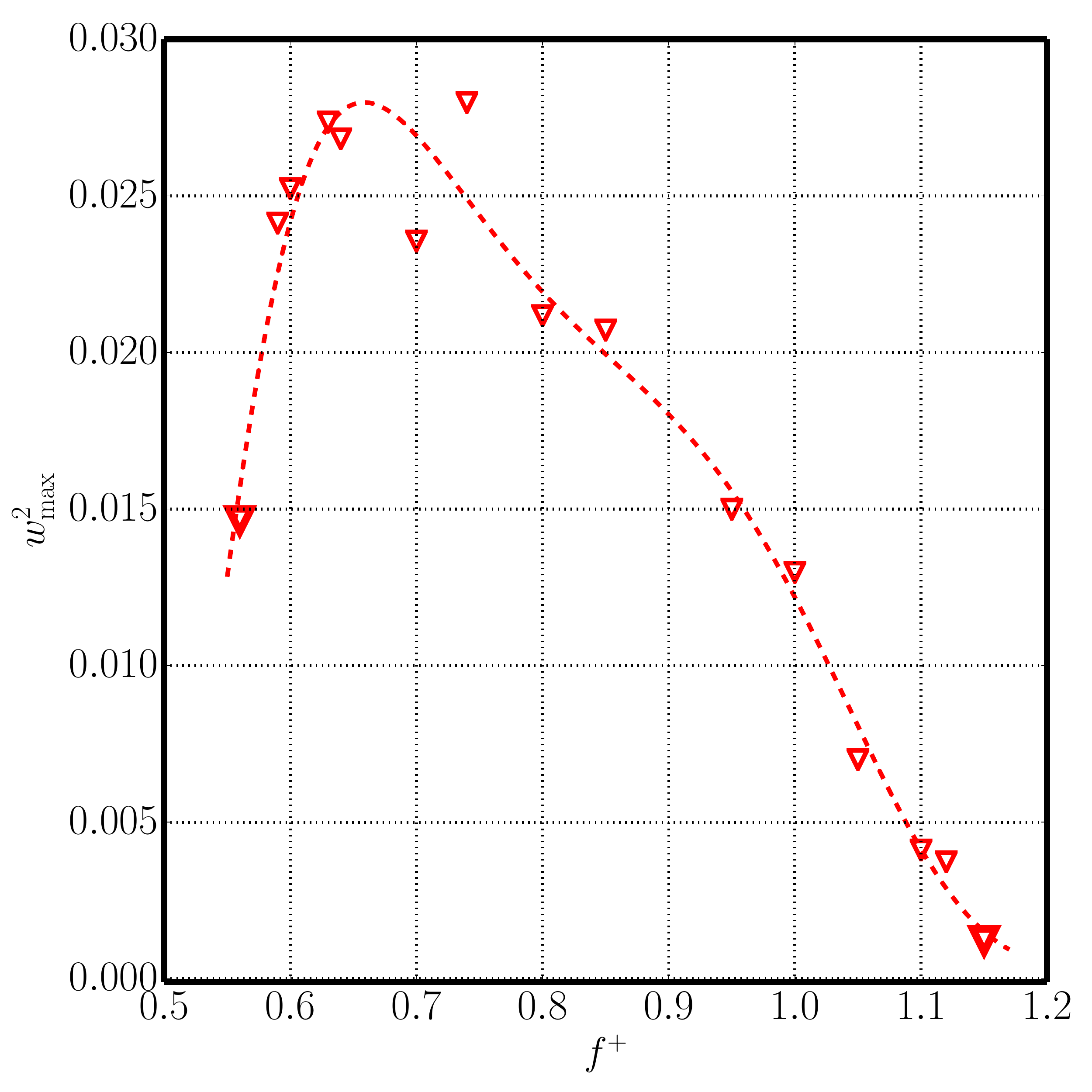}}
\caption{a) Contours of the 
transverse component of the velocity $w$ at a cut plane $y=D/2$. Solid and 
dashed lines correspond to positive and negative contour values respectively. 
b) Maxima of 
$w_{rms}$ for a fixed forcing frequency 
$f^+=0.75$ as a function of $T$ resulting from varying forcing amplitudes. 
c) Maxima of $w_{rms}$ for a constant forcing amplitude $A=4.00$ as a function 
of the forcing frequency $f^+$.}\label{scale_fig1}
                  \end{center}
\end{figure}
Figure \ref{scale_fig1-a} shows 
instantaneous contours of the span-wise velocity $w$ for a plane at $y=D/2$,
revealing Taylor-Couette like vortices, with a wavelength $\lambda_z$ that does
not change with
respect to the forcing amplitude within the range $3<A<4$. We can describe the
flow with a 
Taylor number based on equation (\ref{taylor1}) considering  $\lambda_z$ as a
characteristic length scale and $\omega_i r_i=A U_\infty$. 
We compute the intensity of the velocity fluctuations as $w_{rms}^2=\int_0^{T_f
}
(w-\bar w)^2 dt /T_f$, where $\bar w$ is the time average of $w$ and $T_f$ 
the forcing period $T_f=1/f^+$. The result allows us to identify a maximum value
that 
characterises the intensity of the 3D structure for the forcing case
considered. 

An additional characterisation is possible by studying the amplitude of this
fluctuations  as a 
function of the forcing parameters $(f^+,A)$. We use the 3D DNS to study the
flow 
modifications for two fixed forcing frequencies $f^+=0.75$ and $f^+=1.00$ .  A
useful criterion to quantify the intensity of $3D$ structures is to follow the 
evolution of $w$. Given that the 3D structures are 
present  for $A=4.00$ (case depicted on Figure \ref{resul1}), we decrease the 
forcing amplitude from this value until they vanish. 
In Figure \ref{scale_fig1-c}, the maxima of $w_{rms}$, $w_{\max}$, are plotted 
against the Taylor numbers resulting from equation (\ref{taylor1}), where the
characteristic length scale $\lambda_z$ is found to be $1.16 D$, and the
corresponding forcing 
amplitudes. We can appreciate, looking at the square of the forcing amplitudes,
that the 3D structures become
damped linearly as we approach a threshold at $T=202$ for $f^+=0.75$ and $T=147$
for $f^+=1$. The behaviour
is common to supercritical bifurcations.\\
Another scenario shows up when we follow the evolution of the intensity of 3D 
structures for a fixed forcing amplitude. There is a range of frequencies for 
which the instability develops. This is shown in Figure  \ref{scale_fig1-d}
where $w_{rms}^2$ is observed for $A=4.00$ and
the forcing frequency varying in a range $0.50<f^+<1.20$. We can appreciate that
for 
$f^+\rightarrow1.20$, $w_{\max}^2$ decreases linearly. 
On the other hand, for lower frequencies, we  observe that the 3D instability
appears, with a finite value, 
for $w_{\max}$, at $f^+\ge 0.55$.

\begin{figure}[t]\begin{center}
 \subfigure[]{\label{Taylornumber2-a}
 \includegraphics[height=.49\textwidth]{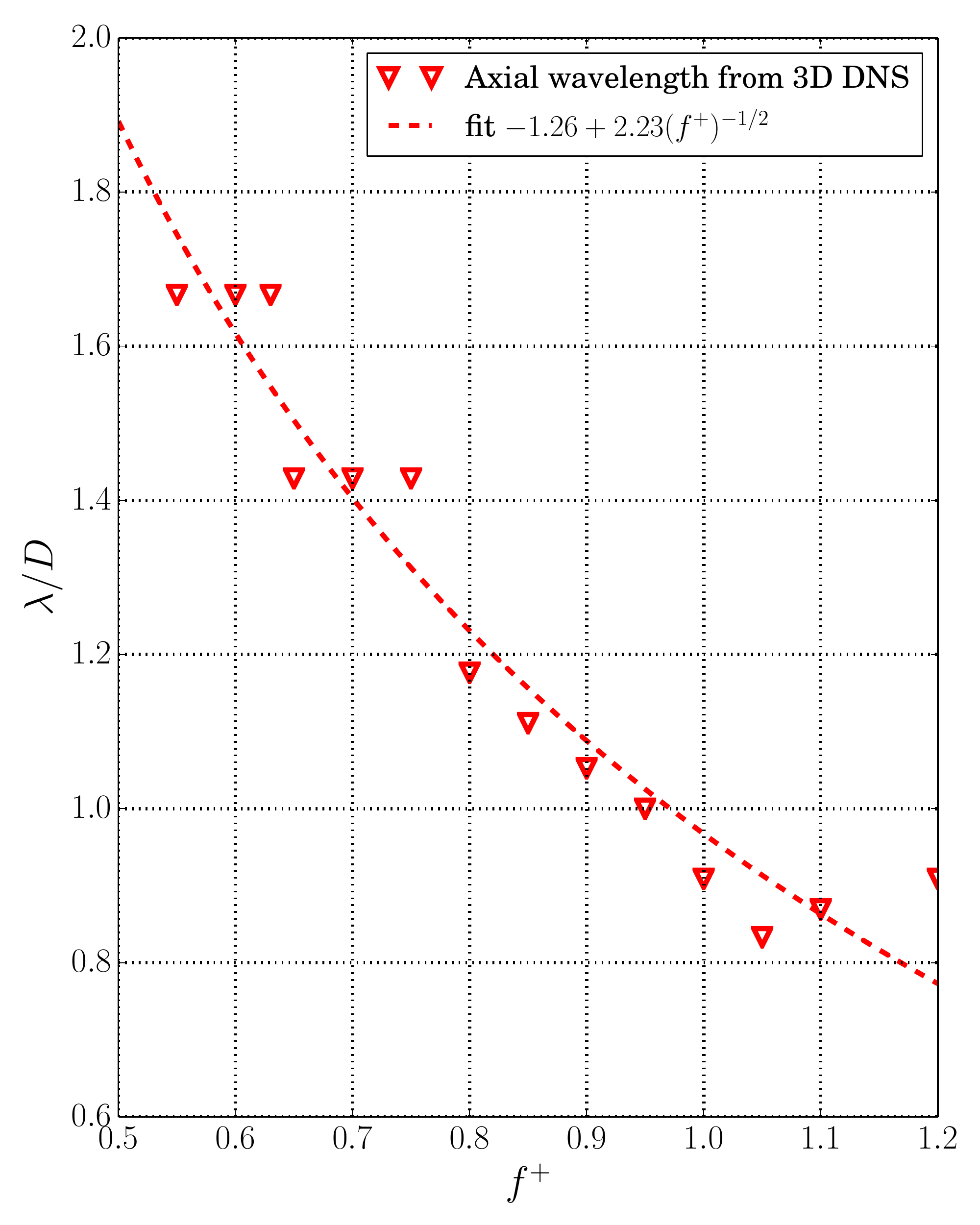}}
 \subfigure[]{\label{Taylornumber2-b}
\includegraphics[height=.49\textwidth]{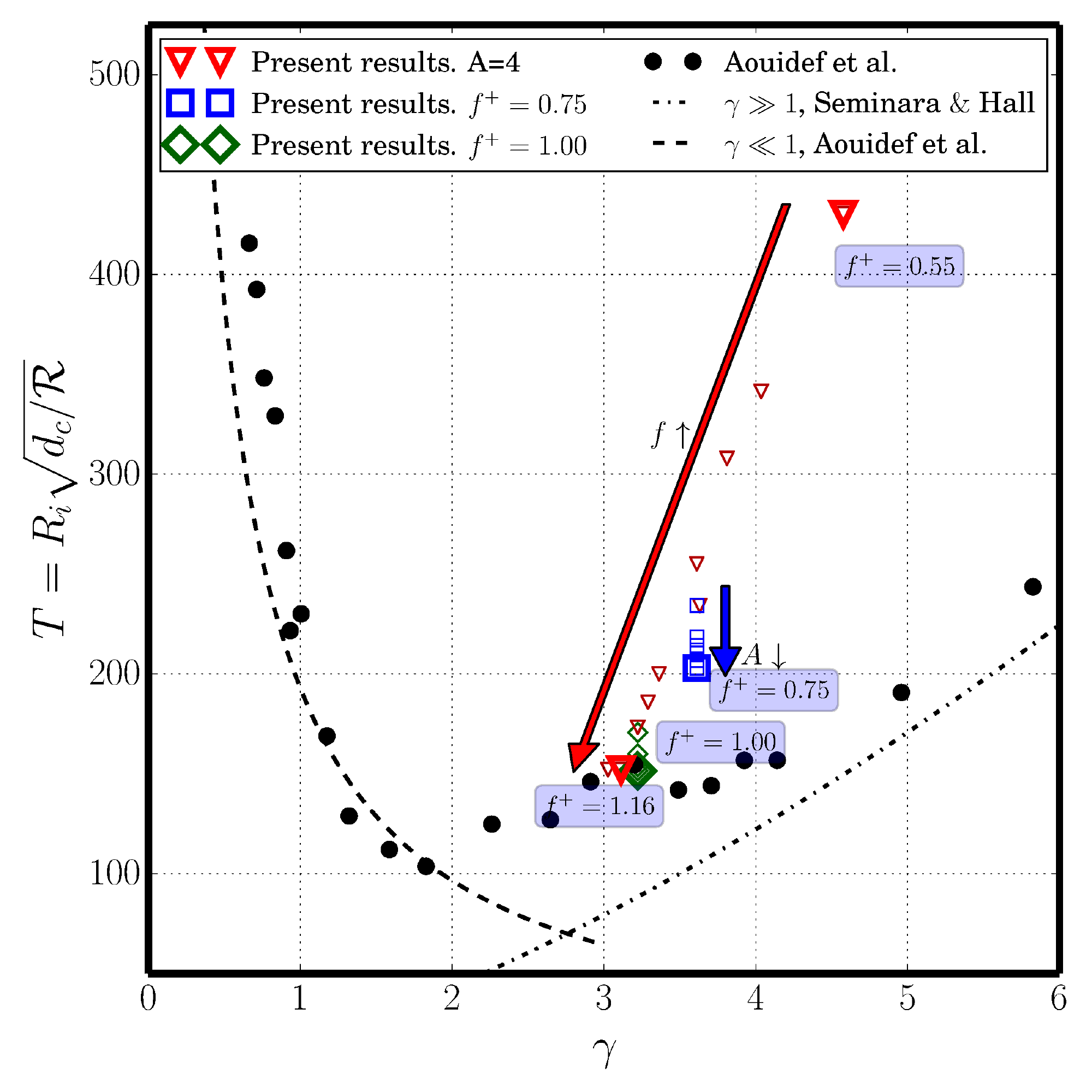}}
\caption{a) For a  fixed forcing amplitude $A=4.00$, corresponding wavelength
measured in  3D simulations.  b) Taylor number versus $\gamma$ estimated from
3D 
simulations. Plotted values correspond to  a fixed forcing amplitude $A=4.00$
and to fixed forcing frequencies $f^+=0.75$ and $f^+=1.00$. Threshold points are 
marked in thicker symbols.  The arrows represent the evolution to 
threshold from Fig. \ref{scale_fig1-c} and Fig. \ref{scale_fig1-d}. Our 
results are compared with experimental and theoretical data that define  
instability threshold in \textit{pure} pulsed flows  from \cite{aouidef1994} and 
\cite{seminara1976}}\label{Taylornumber2}    
              \end{center}
\end{figure}
We perform simulations for different forcing frequencies at a fixed forcing 
amplitude $A=4.00$ in order to characterise the evolution of the wavelength
$\lambda_z$.  We observe in Figure \ref{Taylornumber2-a} that $\lambda_z$ 
depends on $f^+$ following a 
law $\propto (f^+)^{-1/2}$. If we assume that the ``gap'' size $d$ is 
proportional to $\lambda_z$, from the Taylor number definition in equation 
(\ref{taylor1}), where $T$ depends on $d^{3/2}$, then we expect that high
forcing 
frequencies produce decreasing Taylor numbers. This could explain the damping
of 
3D fluctuations for higher frequencies in Figure \ref{scale_fig1-d}. In
addition, we observe that the wavelength
$\lambda_z$  is practically invariant with respect to the amplitude for a given
frequency. 

In studies of pulsed centrifugal instabilities, 
\cite{riley1976,carmi1981,aouidef1994} classified flow regimes 
based on a parameter $\gamma=\sqrt{\omega d_c^2/2\nu}$ which is
the ratio of  a centrifugal region length $d_c$ to the Stokes layer thickness.
In our experiment, $\gamma$  is limited to a range between 2 and 5, it
does not depend on the forcing frequency  and $d_c\sim \lambda$
behaves with respect to $f^+$ as described in Figure \ref{Taylornumber2-a}, 
where 
$\lambda$ decreases almost linearly as $(f^+)^{-1/2}$.

Even though the threshold for centrifugal instabilities determined in the
Taylor-Couette pulsed flow is not directly applicable for a configuration with  
crossflow, the transformation of Taylor numbers based on 
the characteristic length $d_c$ allows an approach for our results. This 
case presents similarity with the eccentric Taylor-Couette instability problem
(see e.g. \cite{leclercq2013,shu2004,shoasiong2006}
and references therein). Indeed, in those problems, the axial wavelength of the
critical perturbations is always of the same order of magnitude of the gap. 

Figure \ref{Taylornumber2-b} summarises 
the stability curves $(\gamma,T)$ for centrifugal pulsed flow determined by 
\cite{aouidef1994,seminara1976} together with the values issued from our  3D
simulations. 
Two analytical curves show the solution 
corresponding to low values of $\gamma$, $T_c = 193.23 \gamma^{-1}$ and high 
values of $\gamma$, $T_c=15.28 \gamma^{3/2}$. The curves are 
supported with experimental data from \cite{aouidef1994}.
On the other hand, within these reference threshold frame, we plotted from our
results 
$T$ against $\gamma$ for a fixed forcing amplitude $A=4.00$, 
and for fixed forcing frequencies $f^+=0.75$ and $f^+=1.00$ (the same data used
to 
construct Figure \ref{scale_fig1}). We observe that the points are contained in
the 
unstable region defined by the analytical curves. For $A=4.00$, the instability
develops 
for $0.55<f^+<1.16$. When $f^+=1.16$, the critical point ($\gamma=3.06$, 
$T=152$) is in very good agreement with the experimental results from pure 
pulsed flows. For decreasing frequencies, $T$ increases almost linearly 
regarding the estimated $\gamma$ until for $f^+=0.55$ the flow stabilises with
respect to centrifugal disturbances ($\gamma=4.4$, 
$T=433$). For a fixed frequency $f^+=0.75$, $\gamma=3.61$, the flow 
destabilises at $T=202$ and with increasing forcing amplitudes, $T$ 
eventually reaches the previous set of points at $A=4.00$. The same behaviour is
found 
for the  fixed forcing frequency $f^+=1.00$, where the flow is unstable from
$T=147$.

\definecolor{green2}{HTML}{4CA54C}
 \sethlcolor{green2}

\begin{figure}[t]
\begin{center}
\includegraphics[width=.65\textwidth]{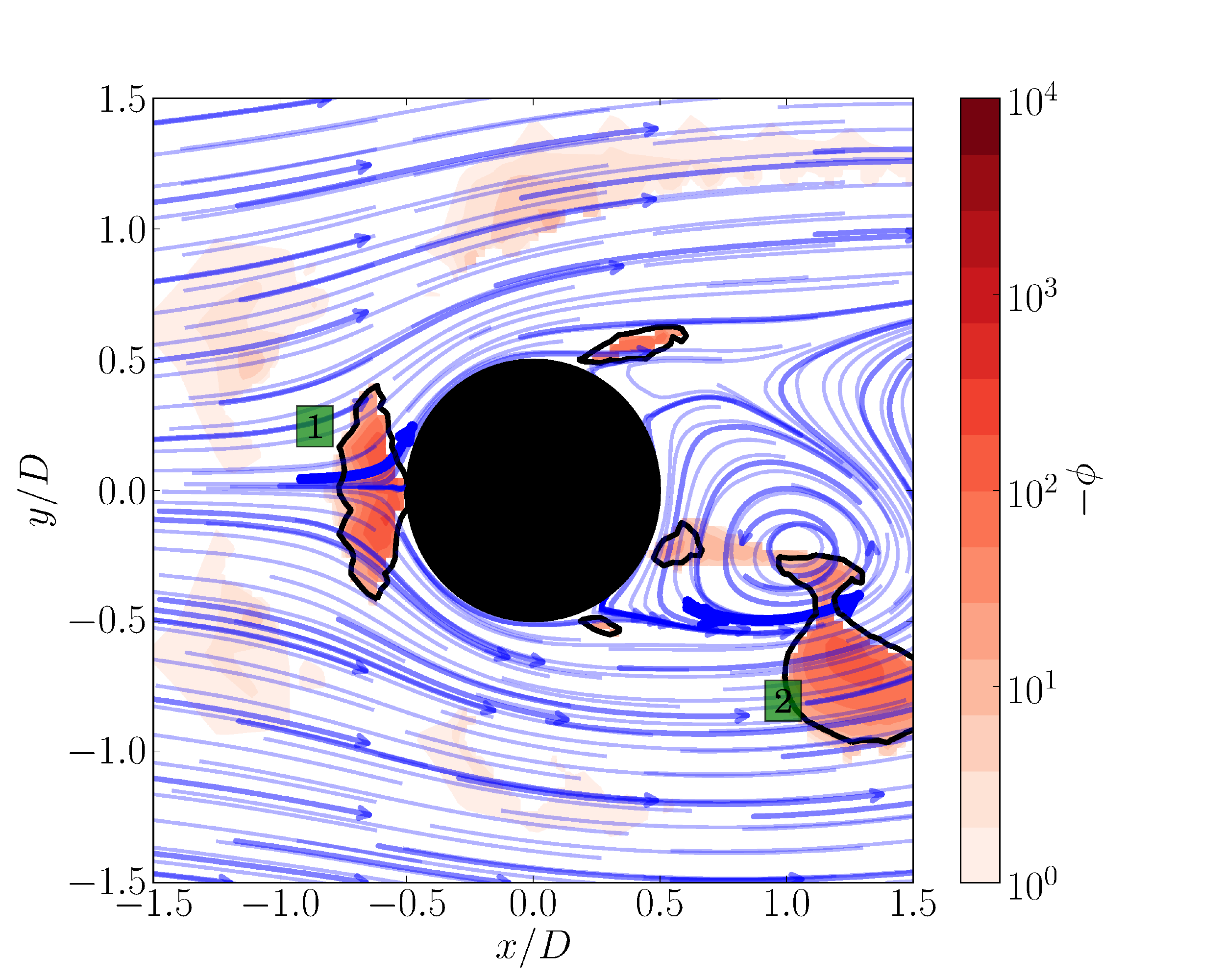}
\caption{In non-forced wake, instantaneous streamlines and $\phi$ negative 
contours reveal potentially centrifugal instabilities regions. Regarding 
curvature and velocity modulus, two distinct regions  associated with 
G\"ortler [{\hl{1}}] and Taylor-Couette [{\hl{2}}] 
mechanisms are marked. However, $-\phi$ is not strong enough to  overcome the 
stabilising effect of the viscosity and to produce 3D instabilities for 
$Re=100$.}\label{teoriaTC-b}     
 \end{center}
\end{figure}
\begin{figure}[h]
\begin{center}
\includegraphics[width=1\textwidth]{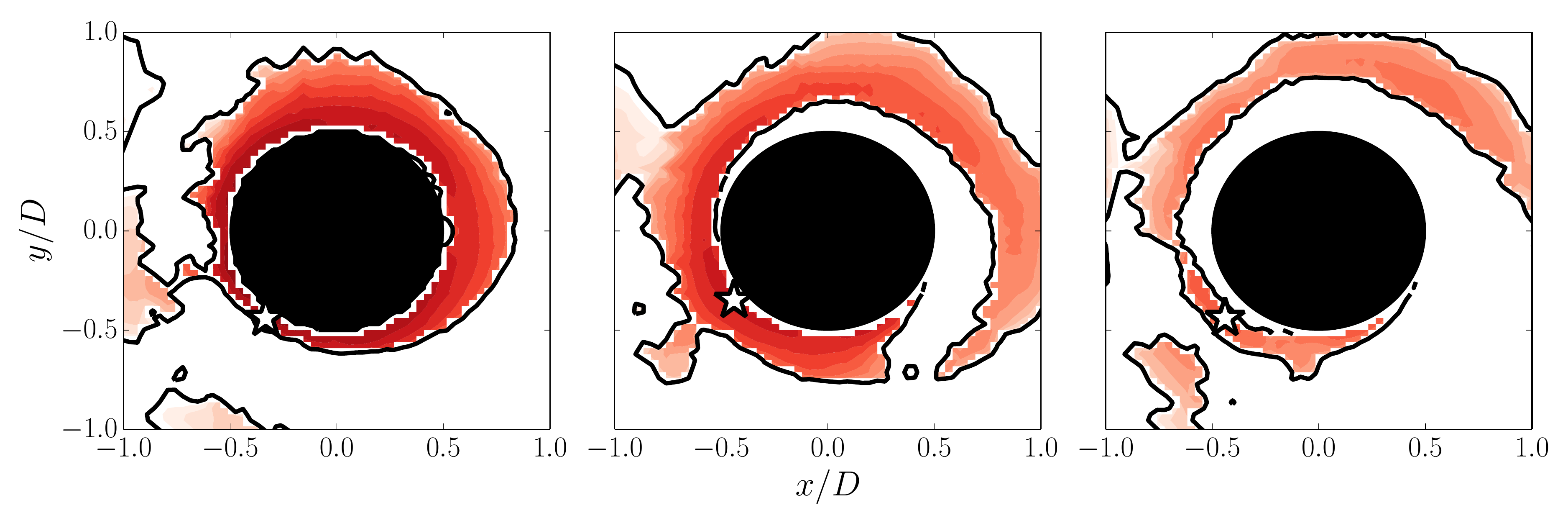}\\
\includegraphics[width=.5\textwidth,trim=0mm 0mm 0mm 190mm,clip]{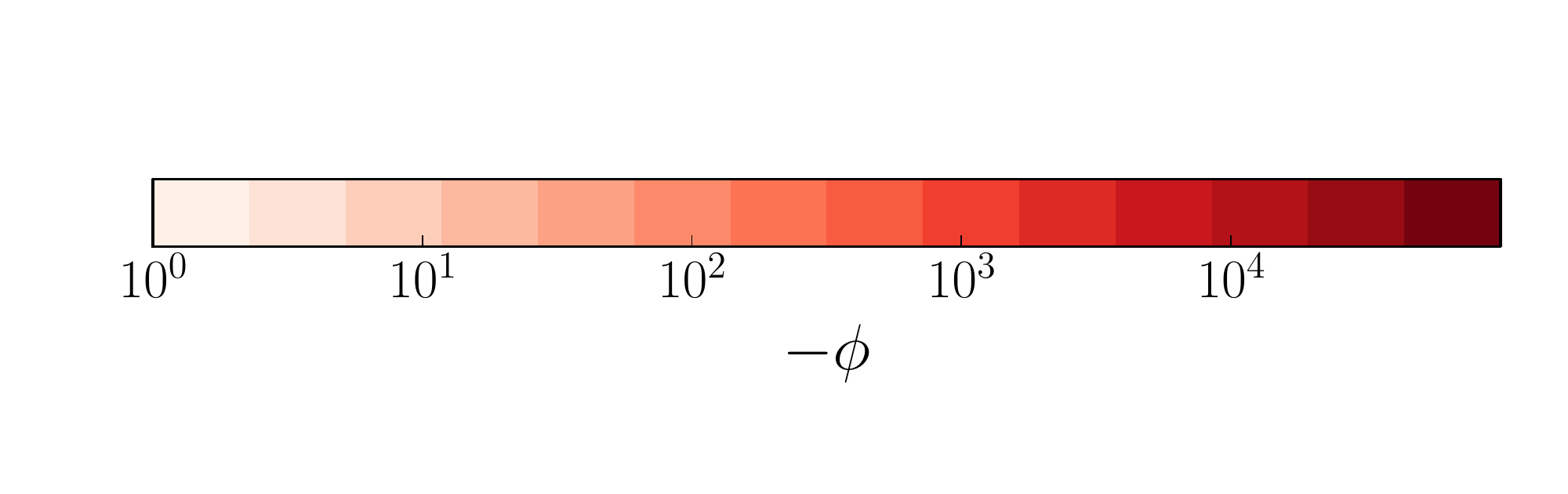}
\caption{Rayleigh discriminant negative values contours, $\phi<0$ for 
$f^+=0.75$ and $A=4.00$, for three different times or oscillation phases within
half of the forcing period $T_f$. The left 
picture shows $\phi$ spatial distribution when a vortex is shed below $x=0$
axis, when the forcing phase $\alpha\simeq 1$ .
Middle ($\alpha\simeq 2$) and right ($\alpha\simeq 3$) pictures allow to 
appreciate the temporal and spatial evolution 
of $\phi$. Symbol  \textcolor{black}{\ding{75}} stands for the position $\bar
x_{\max}$ of $|\phi|_{\max}$. }\label{dc2}       
              \end{center}
\end{figure}

We suggest that the centrifugal instability that develops in the forced wake can
be thus considered in the context of pure 
rotatory pulsed oscillations. Nevertheless, the natural BvK dynamics plays an 
important role as the first bifurcation depends on the distance of the 
forcing state space parameters ($f,A$, Figure \ref{introfig}) to the resonance 
centred at $(f^+=1.00, A=0.00)$. This fact could explain that at $f^+<0.55$ the 
centrifugal instability is not strong enough even when $T$ is high. Conversely, 
for $(f^+=1.16,A=4.00)$ the length $d_c$ is significantly 
smaller and $T$ decreases to the values predicted by the \textit{pure} pulsed flow 
threshold. 

We bring a quantitative picture of these ideas in the remainder of the paper,
starting with a brief review of the criterion for centrifugal instability.

\definecolor{dark-gray}{gray}{0.15}

\section{Centrifugal instability}\label{theory_back}

The necessary condition for a 3D centrifugal instability in flows with curved 
streamlines is given by \cite{rayleigh1916} criterion for inviscid flow \cite[see 
e.g][]{DrazinReid1981} which can be written for flows
such as the Taylor-Couette flow   in terms of the 
Rayleigh discriminant 
\begin{equation} 
 \phi(r)=\frac{2V}{r}\left(\frac{V}{r}+\frac{dV}{dr}\right) \; ,
\end{equation}
\noindent where $V(r)$ is the 2D velocity of an orthoradial base flow field. 
3D perturbations to this flow field are amplified if $\phi(r) < 0$, which 
translates the fact that the perturbed pressure field does not balance the 
centrifugal force, leading to flow instability. 
For a general profile $V(r)$, the flow field can be subdivided in regions of
different stability depending on the sign of $\phi(r)$: it will be unstable in
the region where $\phi(r) < 0$ and
stable when $\phi(r) > 0$. More generally, for other geometries described by a 
vorticity field $\omega_z$, the Rayleigh discriminant can be written as $\phi= 
V(r) \omega_z (r)/\mathcal{R}$  \cite[][]{leblanc1998,sipp2000}, where the local
radius of curvature of the streamlines $\mathcal{R}$ is defined by
 \begin{equation}\label{rayl2}
           \mathcal{R}=\frac{U^2}{(\Delta \psi)(u\nabla u)} \; .
           \end{equation}

\noindent \cite{beaudoin2004} made use of these expressions in order to identify
potential 
instability regions in a 
backward-facing step flow and characterise the 3D global instability. In the 
present case, we will see that the study of the local
Rayleigh discriminant is a useful tool to predict the centrifugal stability of 
the forced cylinder wake problem, which lacks of symmetry simplifications. In 
Figure \ref{teoriaTC-b} the instantaneous flow streamlines along with the 
Rayleigh discriminant $\phi(x,y)$ are represented for a non-forced flow around 
a cylinder at $Re=100$, where the flow produces  the 
B\'enard-von K\'arm\'an vortex shedding. Two distinct regions of potential 
centrifugal instability exist: one near the stagnation point, 
where a concave streamline constitutes a G\"ortler-like geometry
\cite[][]{saric1994}; and another one
in the near wake side, where the curvature of the streamlines around the vortex 
formation region corresponds to a Taylor-Couette 
geometry. At $Re=100$, nevertheless, viscosity prevents the development of 3D
instabilities, which never 
appear for the case shown in Figure \ref{teoriaTC-b}.
When the rotational oscillatory forcing is applied, negative values of $\phi$
appear mostly in regions close to
the cylinder. In what follows we define the characteristic length scale $d_c$ of
regions potentially unstable giving a local Rayleigh criterion to analyse the stability properties of the forced
wake\footnote{It should be noted that viscous effects have to be considered to
determine the actual stability criterion. This could be done using  $d_c$ in the definition of the Taylor number $T$ of equation \ref{taylor1}.}. 

\begin{figure}[t]\begin{center}
\includegraphics[width=.6\textwidth]{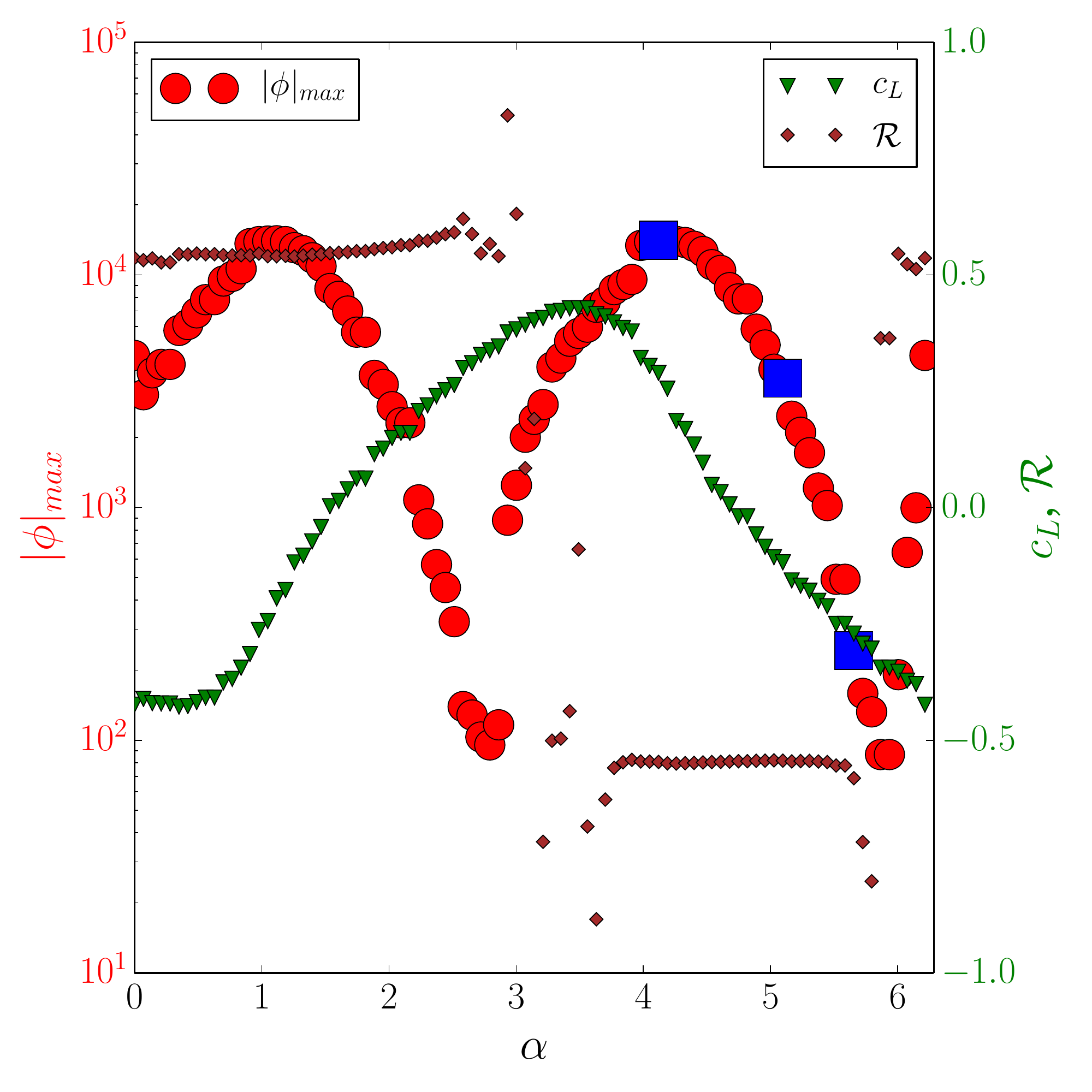}
\caption{$|\phi|_{\max}$, red circle symbols, is the maximum of the Rayleigh discriminant modulus in
function 
of the forcing  phase $\alpha$, $|\phi|_{\max}=|\phi(\bar x_{\max} ,\alpha)|$.
The lift 
coefficient $c_L$, green triangle symbols, gives a reference for the forcing phase $\alpha$. The local
radius  of 
curvature value $\mathcal R$, brown diamond symbols, at $\bar x_{\max}$ is close to the cylinder radius,
its sign 
changes with the shedding cycle. Blue square symbols
\textcolor{blue}{$\blacksquare$} mark
three forcing phases depicted in Fig. \ref{dc2}.}
\label{dc_r}             
              \end{center}
\end{figure}

Despite the flow complexity, it is possible to reduce the problem to 
investigate solely the centrifugal instability of the 2D base flow and its 
relationship with the forcing parameters. We calculate the Rayleigh 
discriminant $\phi(\bar x,t)$ for each forcing parameter from the streamlines 
of the flow at $Re=100$. 
In Figure \ref{dc2}, we present three snapshots of $\phi(\bar x,\alpha)$ at $f^+=0.75$, $A=4.00$ and $Re=100$, where $\alpha \in
[0,2\pi]$ is the forcing phase. For other forcing parameters we obtain the same
qualitative features than what we describe for Figure \ref{dc2}. We observe that a
``corona''-like region appears around the cylinder with negative values of $\phi$. Figure
\ref{dc2}(a) shows the phase when $\phi$ is the most negative, where we can
expect the strongest
possible centrifugal instability with the highest growth rate (\cite{bayly1988}). The
location $\bar x_{\max}$, where the instability can be the most strong locally,  is given
by $ |\phi(\bar x_{\max})|= |\phi_{\max}|$.

Figure \ref{dc2}(b) and \ref{dc2}(c)
describe  the evolution of $\phi(\bar x, \alpha)$ for forcing phases that 
correspond to the mean and the minimum  values, where the flow is less 
receptive to the instability. We observe that $\phi(\bar x, t)$ is
$x$-symmetric 
regarding the forcing phase, $\phi(x,y,\alpha)=\phi(x,-y,\alpha+\pi)$. 
In  Figure \ref{dc_r} we show the variation of the lift 
coefficient $c_L=L /{\frac 1 2 \rho U_{\infty}^2}$, being $L$ the 
resulting lift force, which is
correlated with the phase reference $\alpha$. The local 
radius $\mathcal R$ calculated from Eq. (\ref{rayl2}) in $\bar x_{\max}$ is also
represented, we can see that its modulus is close to the value of the cylinder
radius as the curved streamlines of the forced flow approach the cylinder. The
sign of the local radius accompanies the changes due 
to the oscillation. We present in the same Figure a curve for the evolution of
the maximum of the Rayleigh discriminant modulus
$|\phi(\bar x,t)|_{\max}$ during a forcing period. We indicate over this 
curve the three values of $|\phi|_{\max}$ that lead to the construction of
Figure 
\ref{dc2}. 
\sethlcolor{white}
 \begin{figure}
 \begin{center}
\includegraphics[height=.28\textheight]{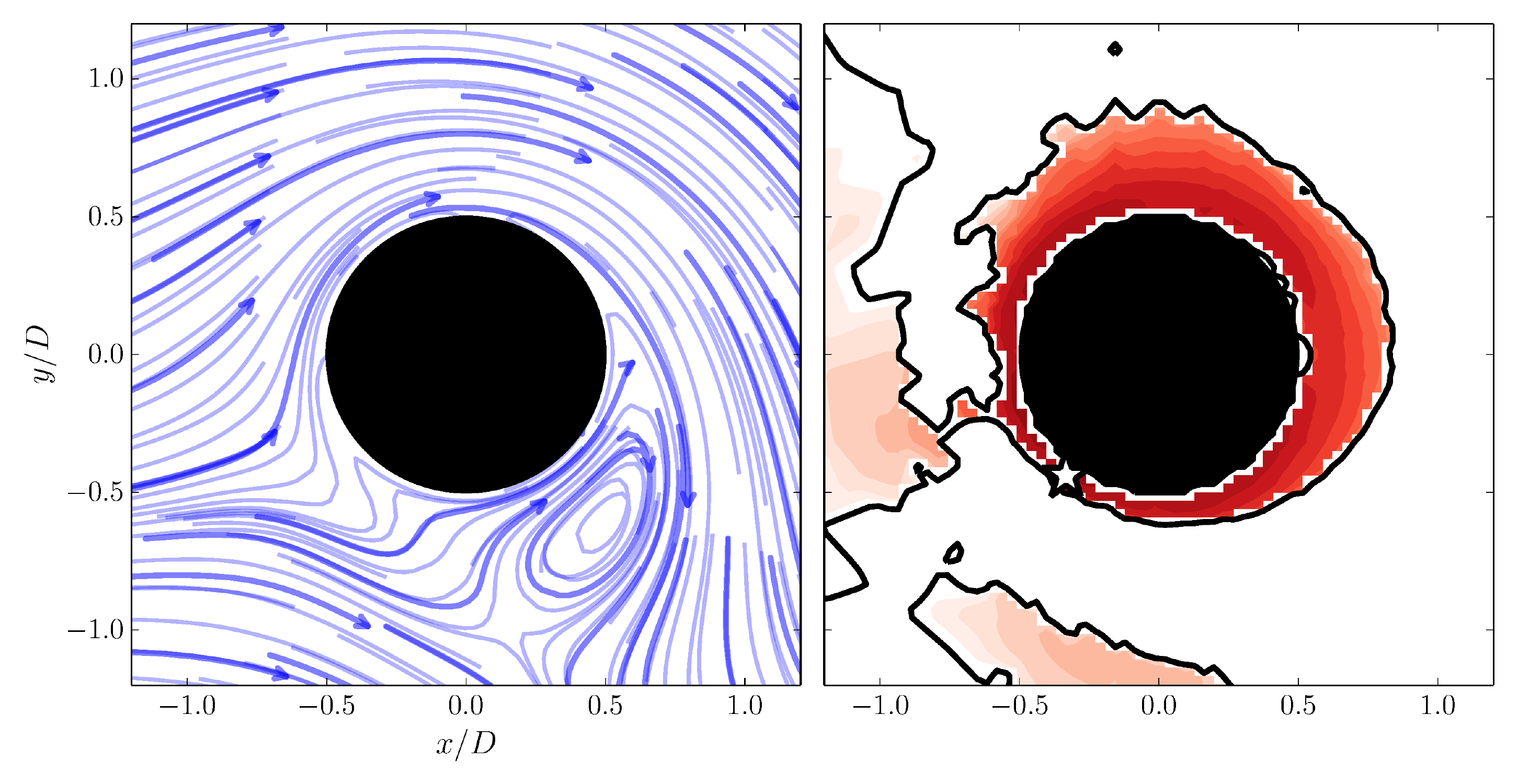}
\includegraphics[height=.28\textheight]{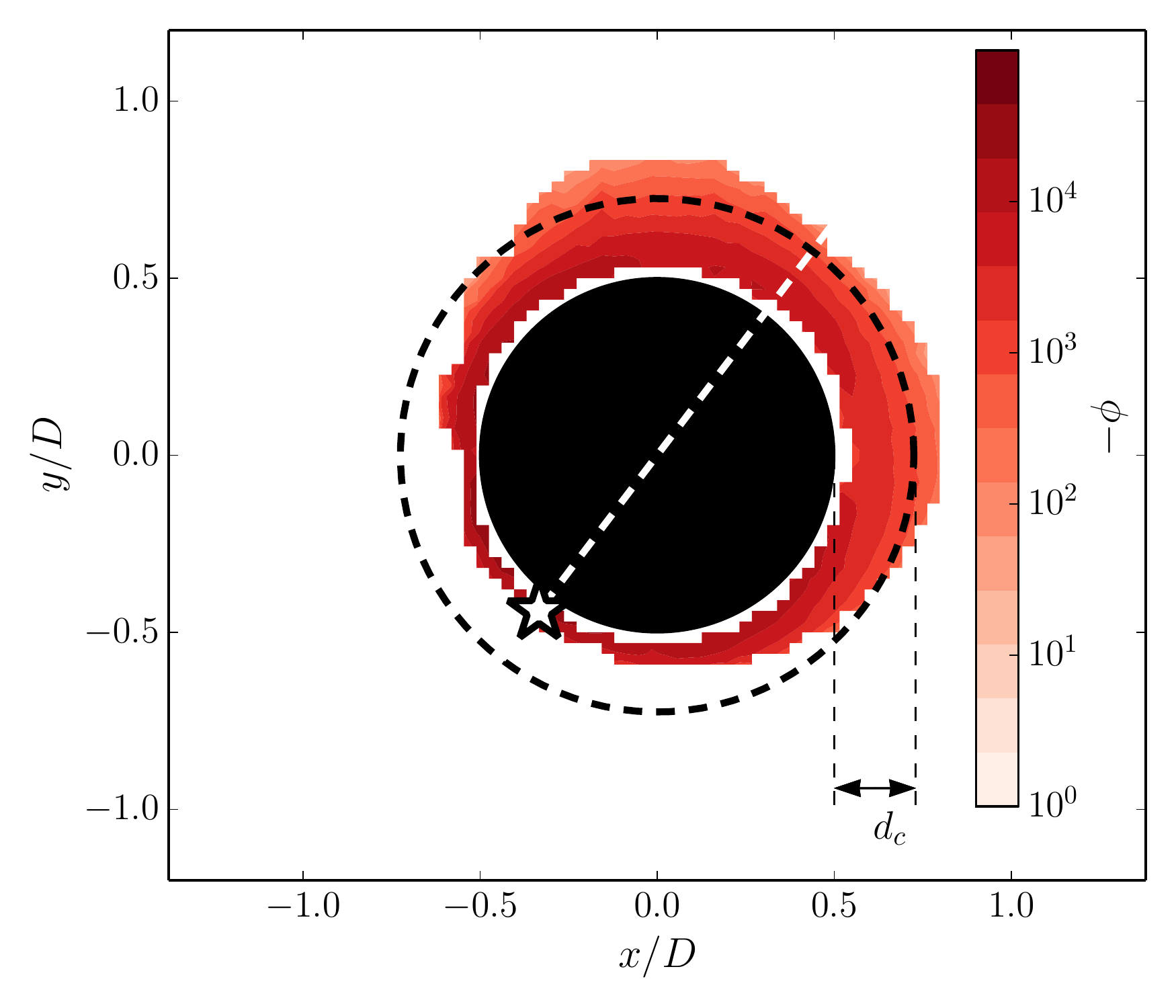}
\put(-170,190){\textcolor{black}{\hl{a)}} }
\put(-30,190){\textcolor{black}{\hl{b)}} }
\put(-80,-5){\textcolor{black}{\hl{c)}} }
 \caption{For $(f^+=0.75,A=4.00)$ at the forcing phase $\alpha_{max}$
corresponding to 
$\phi_{\max}=\phi(\bar x,\alpha_{\max})$: 
a) Streamlines. b) Regions of negative Rayleigh discriminant $-\phi$. The
spatial minimum of $\phi$ is placed at a position
identified with the symbol\textcolor{black}{\ding{75}}. c) $\phi$ regions after
image processing. The corresponding mean radius for
$\phi$ is represented and the resulting measure $d_c$ is indicated.
}\label{dc_final}     
               \end{center}                     
 \end{figure}
 \begin{figure}[t]
 \begin{center}
  \subfigure[]{\label{dc3-a}
 \includegraphics[width=.47\textwidth]{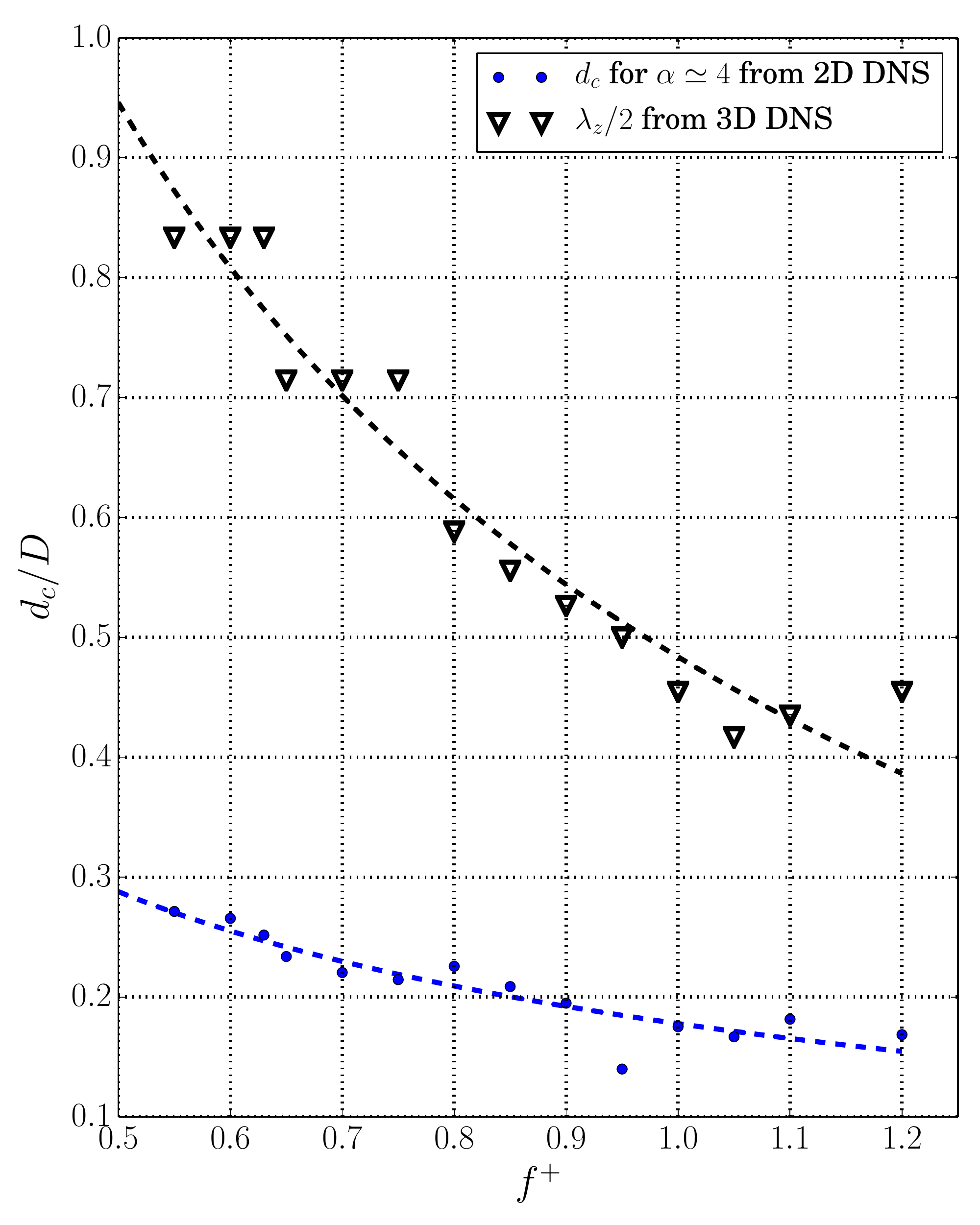}}
   \subfigure[]{\label{dc3-b}
  \includegraphics[width=.47\textwidth]{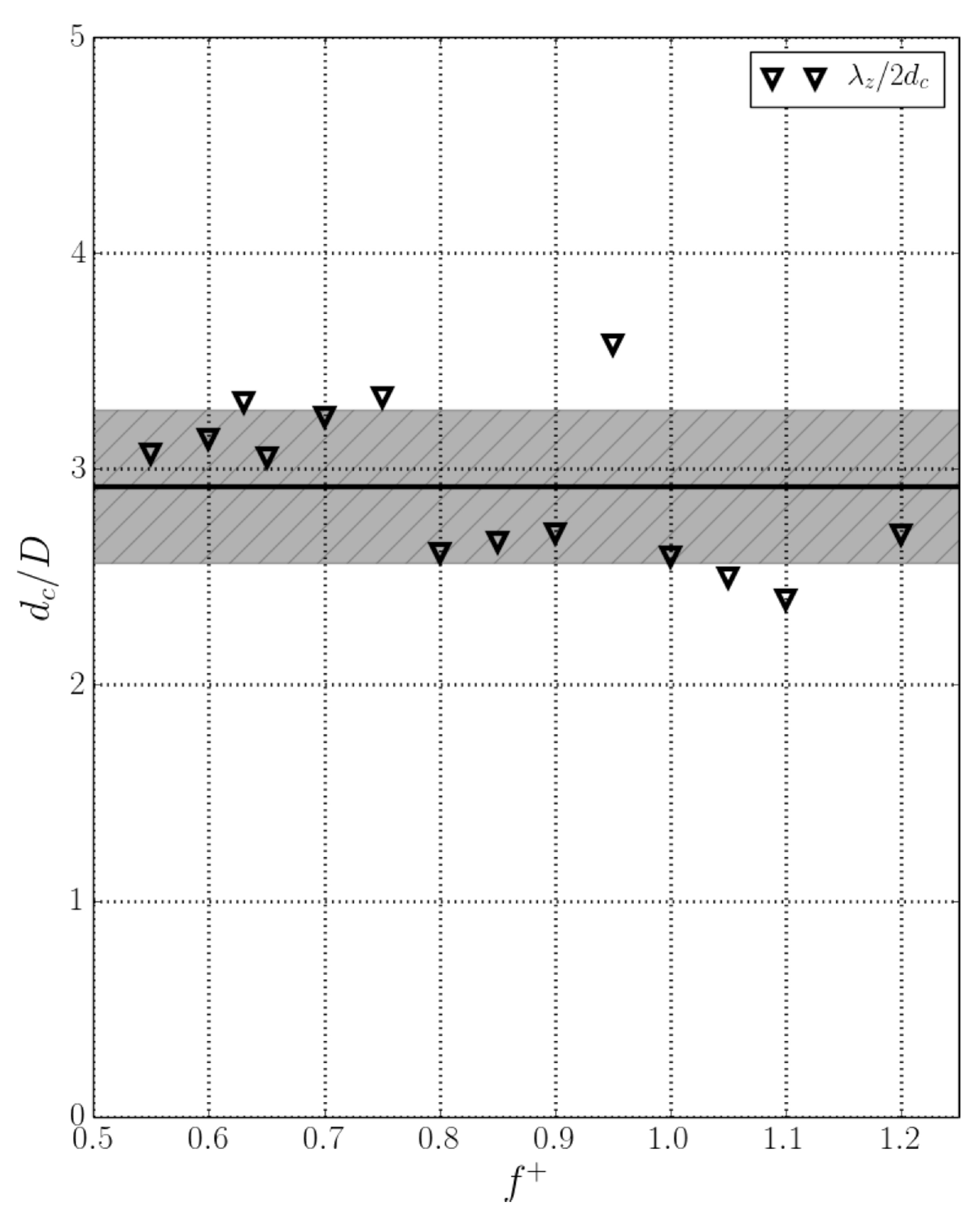}}
 \caption{For a  fixed forcing amplitude $A=4.00$, variation of  the
estimated centrifugal instability measure $d_c$ at $\alpha\simeq 4$ from 2D DNS
and variation of the wavelength $\lambda_z$ associated with the size of the
rolls in Figure \ref{scale_fig1-a}. b) Ratio between the characteristic length
$d_c$ for the centrifugal instability determined from 2D simulations and
$\lambda_z/2$ issued from 3D simulations. Th thick black line represents the 
mean value of $\lambda_z/2$.
}\label{dc3}     
               \end{center}                     
 \end{figure}

As we have already pointed out,  we can 
extract a convenient length scale in order to adapt our problem to the pure
centrifugal instability framework, allowing us to 
compare our results with previous works. If we consider the forcing phase that 
corresponds to $|\phi|_{\max}$, around $\alpha=4$ in Figure \ref{dc_r}, the 2D
 flow streamlines are depicted  by Figure \ref{dc_final}(a). Using equation
\ref{taylor1}, the Rayleigh discriminant is obtained and presented in Figure
\ref{dc_final}(b). 
It is worth mentioning that a $y$-symmetric field is retrieved for $\alpha\simeq
1$ 
that corresponds to the other maximum of  $|\phi|$.  Image processing is used
in 
order to extract a length scale from a contour plot of the Rayleigh
discriminant  
obtained from equation (\ref{taylor1}) as shown in Figure \ref{dc_final}(c) 
(see Appendix for details).  The mean radius represented in Figure
\ref{dc_final}(c) determines the length
scale $d_c$ related to the size of the unstable region for 2D flow. It is shown
in Figure \ref{dc3-a} for different forcing frequencies at a fixed forcing
amplitude A = 4.00 together with the size $\lambda_z/2$ of the centrifugal rolls
that develops in the 3D flow.  Both $d_c$ and $\lambda_z/2$ follow the same
$(f^+)^{-1/2}$ trend, supporting the idea of the pulsed Stokes layer. Their
ratio,  around a value of 3, is plotted in Figure \ref{dc3-b}. Given that
the flow is under non-stationary forcing, the rolls are formed periodically
symmetric with respect to the $x$-axis. Besides, $d_c$ has been determined for a
particular phase $\alpha\simeq 4$, where the instability is most intense, but
the centrifugal instability region changes its size. These arguments may
explain the difficulty for estimating $d_c$  and the the scale 
difference between $\lambda_z/2$ and $d_c$.  Nevertheless, we can observe 
 that the main behaviour is shared between  $\lambda_z$ and $d_c$, therefore, a
centrifugal  instability region observed in 2D simulations is in agreement with
the 3D instability that develops in 3D DNS.

\section{Conclusions}\label{conclusiones}
The present work gives a a new view about 3D instabilities in wake 
flows. In the context of forced wakes at moderate Reynolds numbers, we found a 
new transition that leads to the formation of three dimensional structures. 
The instability shares aspects that were previously studied for centrifugal
pulsed 
flows. Taylor Couette-like vortices  develop from a definite threshold of
forcing parameters
$(f,A)$ and these structures are modified by the incoming flow. For this complex
instability,
2D evaluation of the Rayleigh discriminant $\phi$ may give a fast criterion to 
determine whether a wake flow becomes three dimensional or not. We found from
streamline shapes
and the spatial distribution of $\phi$ that the problem  shares some analogy in
relation to eccentric 
Taylor Couette flows.\\
As 2D forcing in wakes may indeed trigger 3D structures, this behaviour must be 
taken into account in  flow control schemes. Streamlines which result too much
``bent`` by forcing
in wakes can make evident strong negative values of the Rayleigh discriminant
$\phi$ and thus the possibility
of a centrifugal instability.\\
On the other hand, this simple problem can offer an interesting benchmark to
study instabilities 
and transition to turbulence from oscillatory rotation.

\section{Appendix: Determination of  the centrifugal instability region 
 length.}\label{algo_dc}
 The choice of a characteristic length of the centrifugal instability region
from the Rayleigh discriminant scalar fields is not straightforward as we
observe  Fig. \ref{dc2}. we choose to select the forcing phase that corresponds
to the minimum value of $\phi$, the most unstable state. Figure
\ref{dc_final}(b)
presents such state but the $\phi$ scalar field needs to be more clear in order
to extract a length $d_c$. Simple image processing functions,
erosion and dilation, are applied successively to  the scalar field in order to
obtain Fig \ref{dc_final}(c), where a clear shape is noticed. We found
that such shape has an aspect that resembles an eccentric cylinder gap.
Therefore, we choose as a characteristic length the mean radius of this gap
$d_c=1 / (2\pi) \int_0^{2\pi} r d\varphi-D/2$, with $r$ the shape radius varying
with the angular coordinate $\varphi$.

\subsection{Convergence analysis for DNS}
\begin{figure}[p]
 \begin{center}
  \subfigure[]{\label{conv-a}
 \includegraphics[width=.9\textwidth]{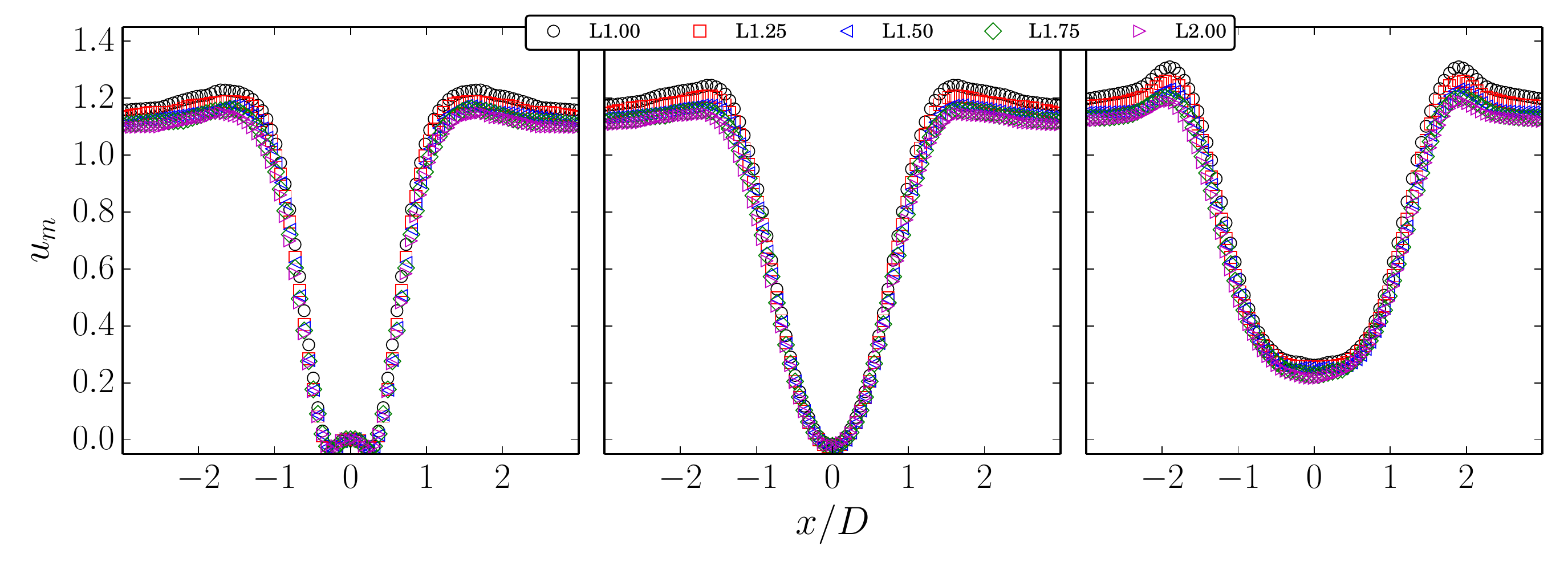}}\\
   \subfigure[]{\label{conv-b}
  \includegraphics[width=.9\textwidth]{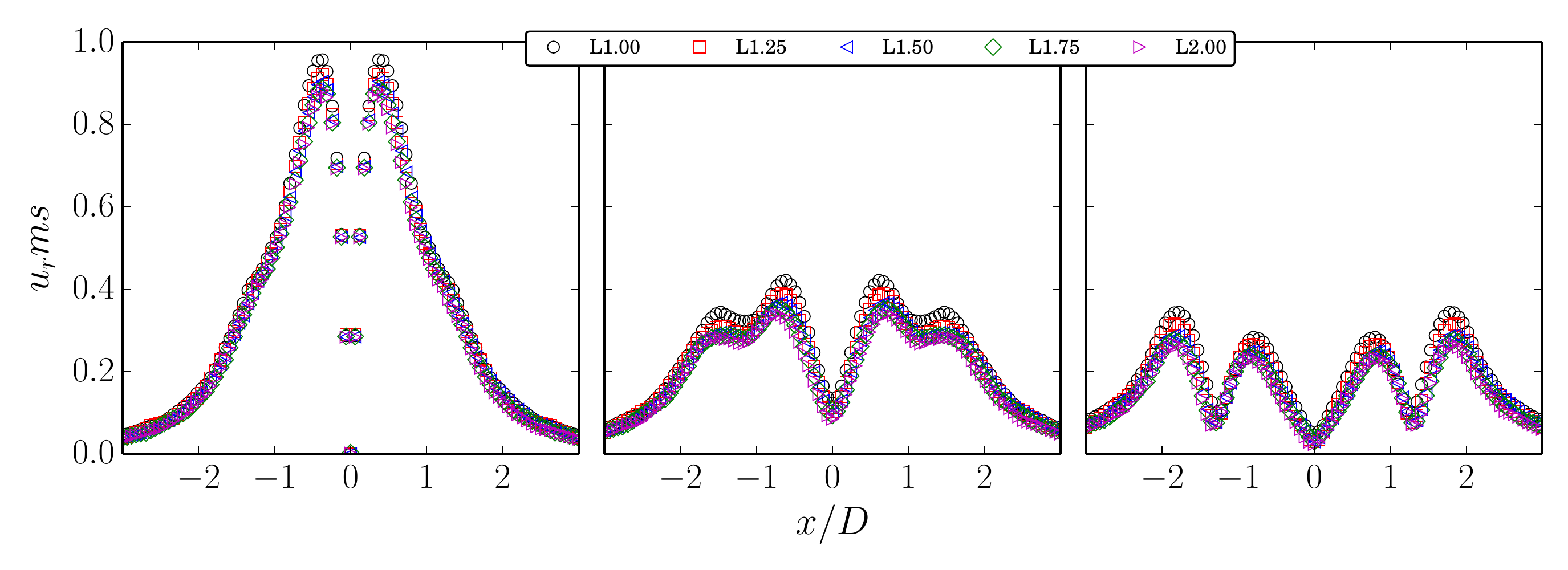}}
 \caption{For the forcing case $f^+=1.00$ and $A=3.50$ we test the reference 
scale $L1.00$ versus larger scales up to $L2.00$ . (a) Mean flow profiles $u_m$ 
at $x=0.5D$, $x=D$ and $x=2D$ for the different scales tested. (b) Fluctuations 
intensity $u_{rms}$ for  flow profiles at the same locations.
}\label{conv}     
               \end{center}                     
 \end{figure}

\begin{figure}
 \begin{center}
 \includegraphics[width=.4\textwidth]{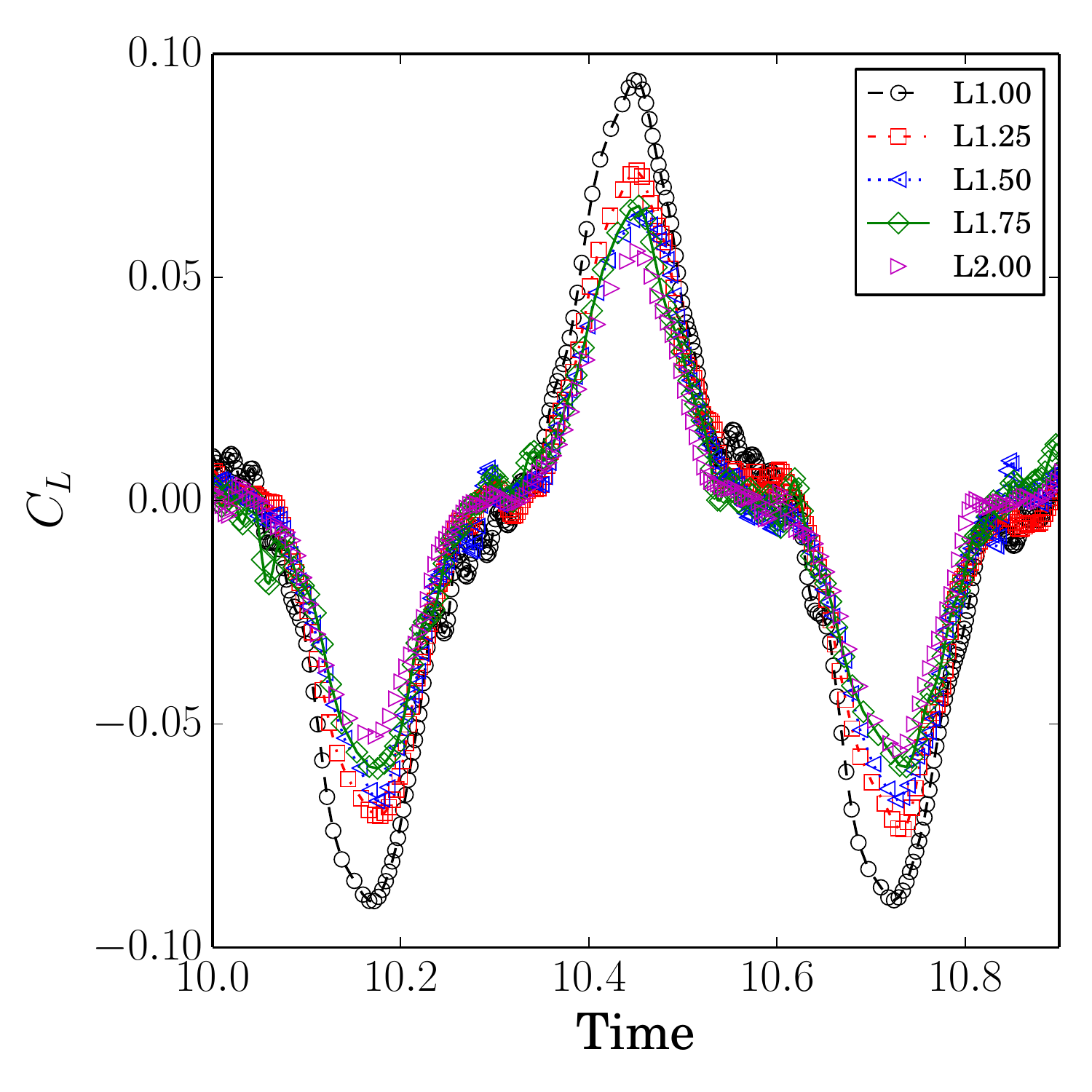}
 \caption{For the forcing case $f^+=1.00$ and $A=3.50$, the lift coefficient 
$C_L$ evolution during two forcing periods corresponding to five different 
scale domains.
}\label{conv2}     
               \end{center}                     
 \end{figure}

In order to ensure that the results do not depend on the size 
of the domain we chose, we performed a convergence analysis for the 2D case.
Given that the  domain size of the reference study is $L_x\times 
L_y=20D\times10D$, we label it as $L1.00$. As we selected larger domains which 
scale as $[1.25;1.50;1.75;2.00]$ the reference study length, we label them 
$L1.25$, $L1.50$, $L1.75$ and $L2.00$. For these scaling lengths,  we plotted 
mean flow profiles for the streamwise component of the velocity $u_m$ at  three 
different $x$ positions $x=0.5D,~D,~2D$   in Fig. \ref{conv-a}.  Figure 
\ref{conv-b} presents for the same direction, fluctuations intensity $u_{rms}$ 
profiles for the same different $x$ positions. We observe that changing the 
domain size does not modify the flow dynamics. Lift coefficient is also 
calculated for each case as it is presented in Fig.\ref{conv2}. We also observe 
good agreement between the different scaling domains.

\section*{Data Accessibility}
 In order to reproduce all the calculations included in this paper, Gerris\citep{Popinet2003572} is available free of charge under the Free Software GPL license and our code files can be downloaded from this  \href{https://drive.google.com/file/d/0BzNBamnN3i2sMnFDaGNpelR5aXM/view?usp=sharing}{URL}.

\section*{Acknowledgment}
We acknowledge support from the LIA PMF-FMF (Franco-Argentinian International Associated Laboratory in the Physics and Mechanics of Fluids), Argentina - France.

\end{document}